\documentclass[12pt]{article}
\usepackage{amsmath}

\usepackage{amssymb}
\usepackage{bm}
\usepackage{braket}
\usepackage[dvips]{graphicx}
\usepackage{comment}

\begin{document}

\baselineskip=17pt

\begin{titlepage}
\rightline{\tt arXiv:2405.10935}
\begin{center}
\vskip 1.5cm
\baselineskip=22pt
{\Large \bf {Nonperturbative correlation functions}}\\
{\Large \bf {from homotopy algebras}}\\
\end{center}
\begin{center}
\vskip 1.0cm
{\large Keisuke Konosu and Yuji Okawa}
\vskip 1.0cm
{\it {Graduate School of Arts and Sciences, The University of Tokyo}}\\
{\it {3-8-1 Komaba, Meguro-ku, Tokyo 153-8902, Japan}}\\
konosu-keisuke@g.ecc.u-tokyo.ac.jp,
okawa@g.ecc.u-tokyo.ac.jp
\vskip 2.0cm

{\bf Abstract}
\end{center}

\noindent
The formula for correlation functions based on quantum $A_\infty$ algebras
in arXiv:2203.05366, arXiv:2305.11634, and arXiv:2305.13103
requires us to divide the action 
into the free part and the interaction part.
We present a new form of the formula which does not involve such division.
The new formula requires us to choose a solution to the equations of motion
which does not have to be real,
and we claim that the formula gives correlation functions
evaluated on the Lefschetz thimble associated with the solution we chose.
Our formula correctly reproduces correlation functions in perturbation theory,
but it can be valid nonperturbatively,
and we present numerical evidence for scalar field theories in zero dimensions
both in the Euclidean case and the Lorentzian case
that correlation functions for finite coupling constants
can be reproduced.
When the theory consists of a single Lefschetz thimble,
our formula gives correlation functions of the theory
by choosing the solution corresponding to the thimble.
When the theory consists of multiple Lefschetz thimbles,
we need to evaluate the ratios of the partition functions for those thimbles
and we describe a method of such evaluations
based on quantum $A_\infty$ algebras in a forthcoming paper.

\end{titlepage}

\tableofcontents

\section{Introduction}\label{introduction}
\setcounter{equation}{0}

String theory has provided us with clues to quantum gravity.
When we explore quantum aspects of string theory
such as mass renormalization and vacuum shift,
we need to go beyond the world-sheet perturbation theory
based on the integration of on-shell vertex operators
over the moduli space of Riemann surfaces.
String field theory provides such a framework.\footnote{
Mass renormalization and vacuum shift were originally discussed
in~\cite{Pius:2013sca, Pius:2014iaa, Pius:2014gza, Sen:2014dqa, Sen:2015hha, Sen:2015uoa},
and the discussion was later reorganized in the review~\cite{deLacroix:2017lif}.
See also subsection~9.3 of the recent review~\cite{Sen:2024nfd} for a concise explanation.}
String field theory may also be useful
when we attempt to construct a framework
where we can prove the AdS/CFT correspondence.
For example, one possible scenario was described in~\cite{Okawa:2020llq}
where it was proposed to evaluate
correlation functions of gauge-invariant operators
for open superstring field theory
in the $1/N$ expansion for that purpose.
While string field theory in the classical theory
has been useful in describing
nonperturbative physics such as tachyon condensation,
we need to study quantum aspects of string field theory
when we explore these problems.

String field theory is a space-time field theory
involving infinitely many fields,
and conceptually it is the same as ordinary field theory to some extent.
However, string field theory is highly complicated
compared to ordinary field theory,
and we need efficient tools to study string field theory.
First of all, constructing a gauge-invariant action of string field theory
can be difficult,
and homotopy algebras such as
$A_\infty$ algebras~\cite{Stasheff:I, Stasheff:II, Getzler-Jones, Markl, Penkava:1994mu, Gaberdiel:1997ia} and $L_\infty$ algebras~\cite{Zwiebach:1992ie, Markl:1997bj}
have played an important role in the construction of gauge-invariant actions.
Recently we have come to recognize that homotopy algebras can also be useful
in studying quantum aspects of string field theory.
An important point is
that the effective action which we obtain when we integrate out part of the degrees of freedom
inherits the structure of the homotopy algebra
such as the $A_\infty$ structure or the $L_\infty$ structure
of the theory before integrating out the degrees of freedom.
Homotopy algebras are thus very useful in describing
the structure of the effective action~\cite{Arvanitakis:2020rrk, Arvanitakis:2021ecw, Sen:2016qap, Erbin:2020eyc, Koyama:2020qfb}.
Furthermore, the description of the effective action in terms of homotopy algebras is universal.
Since the actions of superstring field theory are quite complicated,
we should develop the description in terms of homotopy algebras
in simpler theories before we use it to study quantum aspects of superstring field theory.
With this motivation, we are currently developing technologies of homotopy algebras
using simpler quantum field theories.

As we mentioned before, homotopy algebras are useful
for describing the effective theory when we integrate out part of the degrees of freedom.
When we consider correlation functions in quantum field theory,
we perform the path integral completely,
and this corresponds to integrating out all the degrees of freedom.
This point of view has led us to find a formula
for correlation functions in terms of quantum $A_\infty$ algebras.
A formula for correlation functions for scalar field theories
based on $A_\infty$ algebras was proposed in~\cite{Okawa:2022sjf},
and it was refined to a form which is analogous to string field theory
in~\cite{Konosu:2023pal, Konosu:2023rkm} and extended to incorporate Dirac fields.\footnote{
The description of global symmetries and their anomalies
based on this formula was developed recently in~\cite{Konosu:2024dpo}.
}
The formula is compactly written as
\begin{equation}
\langle \, \Phi^{\otimes n} \, \rangle
= \pi_n \, \frac{1}{{\bf I} +{\bm h} \, {\bm m} +i \hbar \, {\bm h} \, {\bf U}} \, {\bf 1} \,.
\end{equation}
We will explain the ingredients of this formula later,
but it involves an inverse of ${\bf I} +{\bm h} \, {\bm m} +i \hbar \, {\bm h} \, {\bf U}$
which is a linear operator acting on a vector space.
The coupling constants are incorporated in ${\bm m}$, and
it was shown in perturbation theory that correlation functions based on this formula
satisfy the Schwinger-Dyson equations
when the inverse of ${\bf I} +{\bm h} \, {\bm m} +i \hbar \, {\bm h} \, {\bf U}$
is defined by
\begin{equation}
\frac{1}{{\bf I} +{\bm h} \, {\bm m} +i \hbar \, {\bm h} \, {\bf U}}
= {\bf I} +\sum_{n=1}^\infty \, (-1)^n \,
( {\bm h} \, {\bm m} +i \hbar \, {\bm h} \, {\bf U} )^n \,.
\end{equation}
It is possible, however, that the inverse of the operator
${\bf I} +{\bm h} \, {\bm m} +i \hbar \, {\bm h} \, {\bf U}$ exists for finite coupling constants,
and in that case our formula may be regarded as a nonperturbative definition of correlation functions
for finite coupling constants.
In this paper we present evidence that this is indeed the case for scalar field theories in zero dimensions.

Even if the inverse of the operator
${\bf I} +{\bm h} \, {\bm m} +i \hbar \, {\bm h} \, {\bf U}$ exists,
one unsatisfactory aspect of our formula as a nonperturbative definition of correlation functions
is that the construction necessarily involves dividing the action into the free part and the interactions.
In this paper we present a new form of the formula which does not involve such division.
It turns out that the new formula requires us to choose a solution to the equations of motion
which does not have to be real,
and we claim that the formula gives correlation functions
evaluated on the Lefschetz thimble associated with the solution we chose~\cite{Witten:2010cx}.
When the theory consists of a single Lefschetz thimble,
our formula gives correlation functions of the theory
by choosing the solution corresponding to the thimble.
When the theory consists of multiple Lefschetz thimbles,
we need to evaluate the ratios of the partition functions for those thimbles,
and we describe a method of such evaluations
based on quantum $A_\infty$ algebras in a forthcoming paper~\cite{Konosu:2024}.

\section{Scalar field theories in zero dimensions}\label{result-intro}
\setcounter{equation}{0}
In this section, we consider scalar field theories in zero dimensions
and we present evidence that our formula describes nonperturbative correlation functions.\footnote{
We follow the conventions in~\cite{Konosu:2023pal, Konosu:2023rkm}
for the description of scalar field theories in terms of $A_\infty$ algebras.
See~\cite{Konosu:2023pal, Konosu:2023rkm} for further details. 
}
We consider both the Euclidean case and the Lorentzian case.

In the Euclidean case, we denote the zero-dimensional scalar field by $\varphi$
and consider the action $S$ given by
\begin{equation}
S = \frac{1}{2} \, m^2 \, \varphi^2
+\frac{1}{3} \, g \, \varphi^3 +\frac{1}{4} \, \lambda \, \varphi^4 \,,
\label{Euclidean-quartic-action}
\end{equation}
where $m$, $g$, and $\lambda$ are real constants.
In the path integral formalism, the partition function $Z$ is given by
\begin{equation}
Z = \int_{-\infty}^\infty d \varphi \, e^{-\frac{S}{\hbar}} \,,
\end{equation}
and the correlation functions $\langle \, \varphi^n \, \rangle$ are given by
\begin{equation}
\langle \, \varphi^n \, \rangle = \frac{1}{Z} \,
\int_{-\infty}^\infty d \varphi \, \varphi^n \, e^{-\frac{S}{\hbar}} \,.
\end{equation}
In the Lorentzian case, the action corresponding to~\eqref{Euclidean-quartic-action}
differs by a sign and is given by
\begin{equation}
S = {}-\frac{1}{2} \, m^2 \, \varphi^2
-\frac{1}{3} \, g \, \varphi^3 -\frac{1}{4} \, \lambda \, \varphi^4 \,.
\label{Lorentzian-quartic-action}
\end{equation}
The partition function $Z$ is defined by
\begin{equation}
Z = \lim_{\epsilon \to 0} \int_{-\infty}^\infty d \varphi \, e^{\frac{i}{\hbar} S_\epsilon}
\end{equation}
with
\begin{equation}
S_\epsilon = {}-\frac{1}{2} \, ( \, m^2 -i \epsilon \, ) \, \varphi^2
-\frac{1}{3} \, g \, \varphi^3 -\frac{1}{4} \, \lambda \, \varphi^4 \,,
\label{S_epsilon}
\end{equation}
where the constant $\epsilon$ is real and positive.
Correlation functions $\langle \, \varphi^n \, \rangle$ are similarly defined by
\begin{equation}
\langle \, \varphi^n \, \rangle = \frac{1}{Z} \,
\lim_{\epsilon \to 0} \int_{-\infty}^\infty d \varphi \,
\varphi^n \, e^{\frac{i}{\hbar} S_\epsilon} \,.
\label{Lorentzian-correlation-functions}
\end{equation}
As we will see, both the Euclidean case and the Lorentzian case
are described in almost the same way in terms of quantum $A_\infty$ algebras.

In the description in terms of quantum $A_\infty$ algebras,
degrees of freedom are described by a vector space which we call $\mathcal{H}$.
When the theory does not have gauge symmetries,
the vector space $\mathcal{H}$ consists of two vector spaces $\mathcal{H}_1$ and $\mathcal{H}_2$:
\begin{equation}
\mathcal{H} = \mathcal{H}_1 \oplus \mathcal{H}_2 \,.
\end{equation}
In the Batalin-Vilkovisky formalism~\cite{Batalin:1981jr, Batalin:1983ggl, Schwarz:1992nx},
$\mathcal{H}_1$ is for fields
and $\mathcal{H}_2$ is for antifields.
The action is described by $\Phi$ in $\mathcal{H}_1$.
In the case of scalar field theories in zero dimensions,
the vector space $\mathcal{H}_1$ is a one-dimensional vector space,
and we denote its single basis vector by $c$.
We expand $\Phi$ in $\mathcal{H}_1$ as
\begin{equation}
\Phi = \varphi \, c \,,
\end{equation}
where we take the coefficient $\varphi$ to be real
and we identify $\varphi$ with the scalar field that appears in the action.
The vector space $\mathcal{H}_{2}$ is also a one-dimensional vector space,
and we denote its single basis vector by $d$.
The vector space $\mathcal{H}$ is graded by $\mathbb{Z}_2$ degree.
The basis vector $c$ of $\mathcal{H}_1$ is degree even,
and the basis vector $d$ of $\mathcal{H}_2$ is degree odd.
We define the symplectic form $\omega$ which is a linear map from $\mathcal{H} \otimes \mathcal{H}$
to a complex number.
In the case of scalar field theories in zero dimensions,
we define $\omega$ by
\begin{equation}
\biggl(
\begin{array}{cc}
\omega \, ( \, c \,, c \, ) & \omega \, ( \, c \,, d \, ) \\
\omega \, ( \, d \,, c \, ) & \omega \, ( \, d\,, d \, )
\end{array}
\biggr)
= \biggl(
\begin{array}{cc}
0 & 1 \\
{}-1 & 0
\end{array}
\biggr) \,.
\end{equation}

The action described by an $A_\infty$ algebra takes a universal form.
In the Euclidean case, the action $S$ in terms of $\Phi$ in $\mathcal{H}_1$ is given by
\begin{equation}
S = \frac{1}{2} \, \omega \, ( \, \Phi, Q \, \Phi \, )
+\sum_{n=2}^{\infty} \, \frac{1}{n+1} \,
\omega \, ( \, \Phi \,, m_n \, ( \, \Phi \otimes \ldots \otimes \Phi \, ) \, ) \,.
\label{Euclidean-action}
\end{equation}
In the Lorentzian case, the action $S$ is written as
\begin{equation}
S = {}-\frac{1}{2} \, \omega \, ( \, \Phi, Q \, \Phi \, )
-\sum_{n=2}^{\infty} \, \frac{1}{n+1} \,
\omega \, ( \, \Phi \,, m_n \, ( \, \Phi \otimes \ldots \otimes \Phi \, ) \, ) \,.
\label{Lorentzian-action}
\end{equation}
The kinetic term is described by $Q$ which is a linear map from $\mathcal{H}$ to $\mathcal{H}$,
and the cubic interactions are described by $m_2$
which is a linear map from $\mathcal{H} \otimes \mathcal{H}$ to $\mathcal{H}$.
Similarly, the interactions involving $n+1$ fields are described by $m_n$
which is a linear map from $\mathcal{H}^{\otimes n}$ to $\mathcal{H}$,
where
\begin{equation}
\mathcal{H}^{\otimes n}
= \underbrace{\, \mathcal{H} \otimes \mathcal{H} \otimes \ldots \otimes \mathcal{H} \,}_{n} \,.
\end{equation}
To reproduce the action~\eqref{Euclidean-quartic-action} in the Euclidean case
or the action~\eqref{Lorentzian-quartic-action} in the Lorentzian case,
we define $Q$ by
\begin{equation}
Q \, c = m^{2} \, d \,, \qquad Q \, d = 0 \,.
\label{Q-definition}
\end{equation}
We define the action of $m_2$ on $c \otimes c$ to be
\begin{equation}
m_2 \, ( \, c \otimes c \, ) = g \, d \,.
\end{equation}
The operator $m_2$ annihilates any element which involves $d$:
\begin{equation}
m_2 \, ( \, c \otimes d \, ) = 0 \,, \qquad
m_2 \, ( \, d \otimes c \, ) = 0 \,, \qquad
m_2 \, ( \, d \otimes d \, ) = 0 \,.
\end{equation}
Similarly, we define $m_3$ to give a nonvanishing element
only when it acts on $c \otimes c \otimes c$
and is given by
\begin{equation}
m_3 \, ( \, c \otimes c \otimes c \, ) =  \lambda \,  d \,.
\end{equation}
Since the action is quartic, we take $m_n$ to vanish for $n > 3 \,.$
Now it is easy to see that the action~\eqref{Euclidean-quartic-action} in the Euclidean case
and the action~\eqref{Lorentzian-quartic-action} in the Lorentzian case
are reproduced by~\eqref{Euclidean-action} and~\eqref{Lorentzian-action}, respectively.

To describe $A_\infty$ algebras, it is convenient to use the coalgebra representation.\footnote{
The coalgebra representation of $A_\infty$ algebras
is explained in detail, for example, in appendix~A of~\cite{Erler:2015uba}
and in~\cite{Koyama:2020qfb}.
}
In the coalgebra representation, we consider linear operators
acting on the vector space $T \mathcal{H}$ defined by
\begin{equation}
T\mathcal{H} = \mathcal{H}^{\otimes 0} \oplus \mathcal{H}
\oplus \mathcal{H}^{\otimes 2} \oplus \mathcal{H}^{\otimes 3} \oplus \ldots \,,
\end{equation}
where we introduced the vector space $\mathcal{H}^{\otimes 0}$
which is a one-dimensional vector space
given by multiplying a single basis vector {\bf 1} by complex numbers.
The basis vector {\bf 1} satisfies
\begin{equation}
{\bf 1} \otimes \Phi = \Phi \,, \quad \Phi \otimes {\bf 1} = \Phi \,, \quad {\bf 1} \otimes {\bf 1} = {\bf 1}
\end{equation}
for any $\Phi$ in $\mathcal{H}$.
We denote the projection operator onto $\mathcal{H}^{\otimes n}$ by $\pi_n$.

For a map $D_n$ from $\mathcal{H}^{\otimes n}$ to $\mathcal{H}$
with $n = 0, 1, 2, \ldots$,
we define an associated operator ${\bm D}_n$ acting on $T \mathcal{H}$ as follows:
\begin{equation}
\begin{split}
{\bm D}_n \, \pi_m & = 0 \quad \text{for} \quad m < n \,, \\
{\bm D}_n \, \pi_n & = D_n \, \pi_n \,, \\
{\bm D}_n \, \pi_{n+1}
& = ( \, D_n \otimes {\mathbb I} +{\mathbb I} \otimes D_n \, ) \, \pi_{n+1} \,, \\
{\bm D}_n \, \pi_m
& = ( \, D_n \otimes {\mathbb I}^{\otimes (m-n)}
+\sum_{k=1}^{m-n-1} {\mathbb I}^{\otimes k} \otimes D_n \otimes {\mathbb I}^{\otimes (m-n-k)}
+{\mathbb I}^{\otimes (m-n)} \otimes D_n \, ) \, \pi_m \\
& \qquad \text{for} \quad m > n+1 \,.
\end{split}
\label{coderivation-definition}
\end{equation}
Here and in what follows we denote the identity operator on $\mathcal{H}$ by ${\mathbb I}$,
and ${\mathbb I}^{\otimes n}$ is defined by
\begin{equation}
{\mathbb I}^{\otimes n}
= \underbrace{\, {\mathbb I} \otimes {\mathbb I} \otimes \ldots \otimes {\mathbb I} \,}_n \,.
\end{equation}
An operator acting on $T \mathcal{H}$ of the form~\eqref{coderivation-definition}
is called a {\it coderivation}.
A coderivation ${\bm D}$ is characterized in terms of the coproduct $\triangle$ as
\begin{equation}
\triangle \, {\bm D} = ( \, {\bm D} \otimes' {\bf I} +{\bf I} \otimes' {\bm D} \, ) \, \triangle \,,
\label{coderivation-coproduct}
\end{equation}
where ${\bf I}$ is the identity operator on $T \mathcal{H}$.
The form of ${\bm D}$ is uniquely specified by the relation~\eqref{coderivation-coproduct}
and $\pi_1 \, {\bm D}$.
See appendix~A of~\cite{Erler:2015uba} for the definition of the coproduct and further details.
In this paper, we frequently introduce a coderivation ${\bf \Phi}$
associated with an element $\Phi$ of $\mathcal{H}$.
It is defined by
\begin{equation}
\begin{split}
{\bf \Phi} \, {\bf 1} & = \Phi \,, \\
{\bf \Phi} \, \pi_1
& = ( \, \Phi \otimes {\mathbb I} +{\mathbb I} \otimes \Phi \, ) \, \pi_1 \,, \\
{\bf \Phi} \, \pi_m
& = ( \, \Phi \otimes {\mathbb I}^{\otimes m}
+\sum_{k=1}^{m-1} {\mathbb I}^{\otimes k} \otimes \Phi \otimes {\mathbb I}^{\otimes (m-k)}
+{\mathbb I}^{\otimes m} \otimes \Phi \, ) \, \pi_m \quad \text{for} \quad m > 1 \,.
\end{split}
\label{Phi-coderivation}
\end{equation}

We define the coderivation ${\bf Q}$
associated with $Q$
and the coderivation ${\bm m}_n$
associated with $m_n$ for each $n$.
We then define ${\bm m}$ by
\begin{equation}
{\bm m} = \sum_{n=2}^\infty {\bm m}_n \,,
\end{equation}
and we define ${\bf M}$ by
\begin{equation}
{\bf M} = {\bf Q} +{\bm m} \,.
\end{equation}
When we consider gauge theories, the action described by the coderivation ${\bf M}$
is gauge invariant if ${\bf M}$ satisfies
\begin{equation}
{\bf M}^2 = 0 \,.
\end{equation}

When we consider an effective theory in terms of degrees of freedom
described by a subspace of $\mathcal{H}$,
we consider a projection onto the subspace.
Homotopy algebras have turned out to provide useful tools
when we deal with such projections.
We consider projections which commute with $Q$,
and we denote the projection operator by $P$:
\begin{equation}
P^2 = P \,, \qquad P \, Q = Q \, P \,.
\end{equation}
We then promote $P$ on $\mathcal{H}$ to ${\bf P}$ on $T \mathcal{H}$ as follows:
\begin{equation}
\begin{split}
{\bf P} \, \pi_0  & = \pi_0 \,, \\
{\bf P} \, \pi_n & = P^{\otimes n} \, \pi_n
\end{split}
\end{equation}
for $n > 0$, where
\begin{equation}
P^{\otimes n} = \underbrace{ \, P \otimes P \otimes \ldots \otimes P \, }_n \,.
\end{equation}
The operator ${\bf P}$ is a {\it cohomomorphism}.
A cohomomorphism ${\bm F}$ is characterized in terms of the coproduct $\triangle$ as
\begin{equation}
\triangle \, {\bm F} = ( \, {\bm F} \otimes' {\bm F} \, ) \, \triangle \,,
\end{equation}
and it is uniquely specified by this relation and $\pi_1 \, {\bm F}$.
See appendix~A of~\cite{Erler:2015uba} for details.
The operators ${\bf Q}$ and ${\bf P}$ satisfy
\begin{equation}
{\bf P}^2 = {\bf P} \,, \qquad
{\bf Q} \, {\bf P} = {\bf P} \, {\bf Q} \,.
\end{equation}

A key ingredient is an operator $h$ satisfying
\begin{equation}
Q \, h +h \, Q = {\mathbb I}-P \,, \qquad
h \, P = 0 \,, \qquad
P \, h = 0 \,, \qquad
h^2 = 0 \,.
\end{equation}
It is called a {\it contracting homotopy}, and physically it describes
propagators associated with degrees of freedom which are integrated out.
We then promote $h$ on $\mathcal{H}$ to ${\bm h}$ on $T \mathcal{H}$ as follows:
\begin{equation}
\begin{split}
{\bm h} \, \pi_0 & = 0 \,, \\
{\bm h} \, \pi_1 & = h \, \pi_1 \,, \\
{\bm h} \, \pi_2
& = ( \, h \otimes P +{\mathbb I} \otimes h \, ) \, \pi_2 \,, \\
{\bm h} \, \pi_m
& = ( \, h \otimes P^{\otimes (m-1)}
+\sum_{k=1}^{m-2} {\mathbb I}^{\otimes k} \otimes h \otimes P^{\otimes (m-1-k)}
+{\mathbb I}^{\otimes (m-1)} \otimes h \, ) \, \pi_m
\end{split}
\end{equation}
for $m > 2$.
The operator ${\bm h}$ is not a coderivation
and instead satisfies the following relation:
\begin{equation}
\triangle \, {\bm h} = ( \, {\bm h} \otimes' {\bf P} +{\bf I} \otimes' {\bm h} \, ) \, \triangle \,.
\end{equation}
The relations involving $Q$, $P$, and $h$ are promoted
to the following relations:
\begin{equation}
{\bf Q} \, {\bm h} +{\bm h} \, {\bf Q} = {\bf I}-{\bf P} \,, \qquad
{\bm h} \, {\bf P} = 0 \,, \qquad
{\bf P} \, {\bm h} = 0 \,, \qquad
{\bm h}^2 = 0 \,.
\end{equation}

When we consider correlation functions, we perform the path integral {\it completely}.
This corresponds to the case where $P$ vanishes:
\begin{equation}
P = 0 \,.
\end{equation}
The associated operator ${\bf P}$ corresponds to the projection onto $\mathcal{H}^{\otimes 0}$:
\begin{equation}
{\bf P} = \pi_0 \,.
\end{equation}
When $P=0$, the conditions for $h$ are
\begin{equation}
Q \, h +h \, Q = \mathbb{I} \,, \qquad h^2 = 0 \,.
\end{equation}
In the case of scalar field theories in zero dimensions
with $Q$ defined in~\eqref{Q-definition}, $h$ is given by
\begin{equation}
h \, d = \frac{1}{m^2} \, c \,, \qquad h \, c = 0 \,.
\end{equation}
The associated operator $\bm{h}$ is
\begin{equation}
\bm{h} = h \, \pi_1
+\sum_{n=2}^\infty ( \, \mathbb{I}^{\otimes (n-1)} \otimes h \, ) \, \pi_n \,.
\end{equation}

The formula for correlation functions based on quantum $A_{\infty}$ algebras is given by
\begin{equation}
\langle \, \Phi^{\otimes n} \, \rangle
= \pi_n \, \bm{f} \, {\bf 1} \,,
\end{equation}
where
\begin{equation}
\Phi^{\otimes n} = \underbrace{\, \Phi \otimes \Phi \otimes \ldots \otimes \Phi \,}_n
\end{equation}
with $\Phi$ being an element of $\mathcal{H}_1$.
The operator $\bm{f}$ in the Euclidean case is
\begin{equation}
\bm{f}
= \frac{1}{{\bf I} +{\bm h} \, {\bm m} -\hbar \, {\bm h} \, {\bf U}} \,,
\end{equation}
and $\bm{f}$ in the Lorentzian case is
\begin{equation}
\bm{f}
= \frac{1}{{\bf I} +{\bm h} \, {\bm m} +i \hbar \, {\bm h} \, {\bf U}} \,.
\end{equation}
In the case of scalar field theories in zero dimensions, the operator ${\bf U}$ is given by
\begin{equation}
{\bf U} = {\bm c} \, {\bm d} \,,
\end{equation}
where ${\bm c}$ and ${\bm d}$ are coderivations associated with $c$ and $d$, respectively.
Recall that the coderivation associated with an element in $\mathcal{H}$
is defined in~\eqref{Phi-coderivation}.
The notation $\langle \, \Phi^{\otimes n} \, \rangle$ should be understood
in terms of the expansion
\begin{equation}
\Phi = \varphi \, c
\end{equation}
as
\begin{equation}
\langle \, \Phi^{\otimes n} \, \rangle
= \langle \, \underbrace{\, \Phi \otimes \Phi \otimes \ldots \otimes \Phi \,}_n \, \rangle
= \langle \, \varphi^n \, \rangle \,
\underbrace{\, c \otimes c \otimes \ldots \otimes c \,}_n \,. 
\end{equation}
Therefore, the correlation functions are given as coefficients
in front of basis vectors of $T\mathcal{H}$.

The operator $\bm{f}$
is a linear map from $T\mathcal{H}_1$ to $T\mathcal{H}_1$,
where the vector space $T \mathcal{H}_1$ is defined by
\begin{equation}
T\mathcal{H}_{1} = \mathcal{H}^{\otimes 0} \oplus \mathcal{H}_{1} \oplus \mathcal{H}_{1}^{\otimes 2} \oplus \mathcal{H}_{1}^{\otimes 3} \oplus \ldots
\end{equation}
with
\begin{equation}
\mathcal{H}_{1}^{\otimes n}
= \underbrace{\, \mathcal{H}_{1}\otimes\mathcal{H}_{1} \otimes \ldots \otimes \mathcal{H}_{1} \,}_{n}
\end{equation}
for $n > 0$.
For scalar field theories in zero dimensions,
an element $\bm{v}$ of $T\mathcal{H}_1$ can be expanded as
\begin{equation}
\bm{v} = \bm{v}_0 \, {\bf 1} +\bm{v}_1 \, c
+\bm{v}_2 \, c \otimes c +\bm{v}_3 \, c \otimes c \otimes c + \ldots \,,
\end{equation}
and we represent $\bm{v}$ as
\begin{equation}
\bm{v} = \left(
\begin{array}{c}
\bm{v}_0 \\
\bm{v}_1 \\
\bm{v}_2 \\
\bm{v}_3 \\
\vdots
\end{array}
\right) \,.
\end{equation}
A linear map ${\bf A}$ from $T\mathcal{H}_1$ to $T\mathcal{H}_1$ with
\begin{equation}
\begin{split}
{\bf A} \, {\bf 1} & = {\bf A}_{00} \, {\bf 1} +{\bf A}_{01} \, c
+{\bf A}_{02} \, c \otimes c +{\bf A}_{03} \, c \otimes c \otimes c + \ldots \,, \\
{\bf A} \, c & = {\bf A}_{10} \, {\bf 1} +{\bf A}_{11} \, c
+{\bf A}_{12} \, c \otimes c +{\bf A}_{13} \, c \otimes c \otimes c + \ldots \,, \\
{\bf A} \, c \otimes c & = {\bf A}_{20} \, {\bf 1} +{\bf A}_{21} \, c
+{\bf A}_{22} \, c \otimes c +{\bf A}_{23} \, c \otimes c \otimes c + \ldots \,, \\
{\bf A} \, c \otimes c \otimes c & = {\bf A}_{30} \, {\bf 1} +{\bf A}_{31} \, c
+{\bf A}_{32} \, c \otimes c +{\bf A}_{33} \, c \otimes c \otimes c + \ldots \,, \\
& ~~\vdots \label{rep-mat}
\end{split}
\end{equation}
is represented in the matrix form as
\begin{equation}
\bf{A} = \left(
\begin{array}{ccccc}
\bf{A}_{00} & \bf{A}_{01} & \bf{A}_{02} & \bf{A}_{03} & \ldots \\
\bf{A}_{10} & \bf{A}_{11} & \bf{A}_{12} & \bf{A}_{13} & \ldots \\
\bf{A}_{20} & \bf{A}_{21} & \bf{A}_{22} & \bf{A}_{23} & \ldots \\
\bf{A}_{30} & \bf{A}_{31} & \bf{A}_{32} & \bf{A}_{33} & \ldots \\
\vdots & \vdots & \vdots & \vdots & \ddots
\end{array}
\right) \,.
\end{equation}
Therefore, the formula can be expressed as
\begin{equation}
\langle \, \varphi^n \, \rangle
= \bm{f}_{n0} \,.
\end{equation}

Let us consider the ingredients of ${\bm f}$ in the matrix form.
The components ${\bf I}_{\, ij}$ of the identity operator ${\bf I}$ are given by
\begin{equation}
{\bf I}_{\, ij} = \delta_{ij} \,.
\end{equation}
In the matrix form, the identity operator ${\bf I}$ is expressed as follows:
\begin{equation}
\bf{I} = \left(
\begin{array}{ccccccc}
1 & 0 & 0 & 0 & 0 & 0 & \ldots \\
0 & 1 & 0 & 0 & 0 & 0 & \ldots \\
0 & 0 & 1 & 0 & 0 & 0 & \ldots \\
0 & 0 & 0 & 1 & 0 & 0 & \ldots \\
0 & 0 & 0 & 0 & 1 & 0 & \ldots \\
0 & 0 & 0 & 0 & 0 & 1 & \ldots \\
\vdots & \vdots & \vdots & \vdots & \vdots & \vdots & \ddots
\end{array}
\right) \,.
\end{equation}

Let us next consider ${\bm h} \, {\bm m}_2$ and ${\bm h} \, {\bm m}_3$. 
Since
\begin{equation}
\begin{split}
{\bm h} \, {\bm m}_2 \, {\bf 1} & = 0 \,, \\
{\bm h} \, {\bm m}_2 \, c & = 0 \,, \\
{\bm h} \, {\bm m}_2 \,
( \, \underbrace{\, c \otimes \ldots \otimes c \,}_j \, )
& = \underbrace{\, c \otimes \ldots \otimes c \,}_{j-2}
\otimes \, h \, m_2 \, ( \, c \otimes c \, )
= \frac{g}{m^2} \, \underbrace{\, c \otimes \ldots \otimes c \,}_{j-1}
\end{split}
\end{equation}
for $j \geq 2$, we have
\begin{equation}
( \, {\bm h} \, {\bm m}_2 \, )_{i0} = 0 \,, \qquad
( \, {\bm h} \, {\bm m}_2 \, )_{i1} = 0 \,, \qquad
( \, {\bm h} \, {\bm m}_2 \, )_{ij} = \frac{g}{m^2} \, \delta_{i, \, j-1} \quad
\text{for} \quad j \ge 2 \,.
\end{equation}
Since
\begin{equation}
\begin{split}
{\bm h} \, {\bm m}_3 \, {\bf 1} & = 0 \,, \\
{\bm h} \, {\bm m}_3 \, c & = 0 \,, \\
{\bm h} \, {\bm m}_3 \, ( \, c \otimes c \, ) & = 0 \,, \\
{\bm h} \, {\bm m}_3 \,
( \, \underbrace{\, c \otimes \ldots \otimes c \,}_j \, )
& = \underbrace{\, c \otimes \ldots \otimes c \,}_{j-3}
\otimes \, h \, m_3 \, ( \, c \otimes c \otimes c \, )
= \frac{\lambda}{m^2} \, \underbrace{\, c \otimes \ldots \otimes c \,}_{j-2}
\end{split}
\end{equation}
for $j \geq 3$, we have
\begin{equation}
( \, {\bm h} \, {\bm m}_3 \, )_{i0} = 0 \,, \quad
( \, {\bm h} \, {\bm m}_3 \, )_{i1} = 0 \,, \quad
( \, {\bm h} \, {\bm m}_3 \, )_{i2} = 0 \,, \quad
( \, {\bm h} \, {\bm m}_3 \, )_{ij} = \frac{\lambda}{m^2} \, \delta_{i, \, j-2} \quad
\text{for} \quad j \ge 3 \,.
\end{equation}
In the matrix form, they are given by
\begin{equation}
{\bm h} \, {\bm m}_2 = \frac{g}{m^2} \, \left(
\begin{array}{ccccccc}
0 & 0 & 0 & 0 & 0 & 0 & \ldots \\
0 & 0 & 1 & 0 & 0 & 0 & \ldots \\
0 & 0 & 0 & 1 & 0 & 0 & \ldots \\
0 & 0 & 0 & 0 & 1 & 0 & \ldots \\
0 & 0 & 0 & 0 & 0 & 1 & \ldots \\
0 & 0 & 0 & 0 & 0 & 0 & \ldots \\
\vdots & \vdots & \vdots & \vdots & \vdots & \vdots & \ddots
\end{array}
\right) \,, \quad
{\bm h} \, {\bm m}_3 = \frac{\lambda}{m^2} \, \left(
\begin{array}{ccccccc}
0 & 0 & 0 & 0 & 0 & 0 & \ldots \\
0 & 0 & 0 & 1 & 0 & 0 & \ldots \\
0 & 0 & 0 & 0 & 1 & 0 & \ldots \\
0 & 0 & 0 & 0 & 0 & 1 & \ldots \\
0 & 0 & 0 & 0 & 0 & 0 & \ldots \\
0 & 0 & 0 & 0 & 0 & 0 & \ldots \\
\vdots & \vdots & \vdots & \vdots & \vdots & \vdots & \ddots
\end{array}
\right) \,.
\end{equation}

The last ingredient is ${\bm h} \, {\bf U}$. Since
\begin{equation}
\pi_{j+2} \, {\bm h} \, {\bf U} \, \pi_j
= \sum_{k=0}^j \, \mathbb{I}^{\otimes k} \otimes c \otimes \mathbb{I}^{\otimes (j-k)} \otimes hd
= \frac{1}{m^2} \sum_{k=0}^j \,
\mathbb{I}^{\otimes k} \otimes c \otimes \mathbb{I}^{\otimes (j-k)}\otimes c \,,
\end{equation}
we have
\begin{equation}
{\bm h} \, {\bf U} \,
( \, \underbrace{\, c \otimes \ldots \otimes c \,}_j \, )
= \frac{j+1}{m^2} \, \underbrace{\, c \otimes \ldots \otimes c \,}_{j+2}
\end{equation}
and
\begin{equation}
( \, {\bm h} \, {\bf U} \, )_{ij} = \frac{j+1}{m^2} \, \delta_{i, \, j+2} \,.
\end{equation}
In the matrix form, it is given by
 \begin{equation}
{\bm h} \, {\bf U} = \frac{1}{m^2} \, \left(
\begin{array}{ccccccc}
0 & 0 & 0 & 0 & 0 & 0 & \ldots \\
0 & 0 & 0 & 0 & 0 & 0 & \ldots \\
1 & 0 & 0 & 0 & 0 & 0 & \ldots \\
0 & 2 & 0 & 0 & 0 & 0 & \ldots \\
0 & 0 & 3 & 0 & 0 & 0 & \ldots \\
0 & 0 & 0 & 4 & 0 & 0 & \ldots \\
\vdots & \vdots & \vdots & \vdots & \vdots & \vdots & \ddots
\end{array}
\right) \,.
\end{equation}

\subsection{The Euclidean case}

We set $g=0$ and consider $\varphi^4$ theory in zero dimensions.
Let us begin with the Euclidean case.
We first verify that $\bm{f}_{n0}$ reproduces $\langle \, \varphi^n \, \rangle$
in perturbation theory.
The action $S$ is given by
\begin{equation}
S = \frac{1}{2} \, m^2 \, \varphi^2 +\frac{1}{4} \, \lambda \, \varphi^4 \,,
\end{equation}
and we set $m^2 = 1$ and $\hbar = 1$.
Since
\begin{equation}
\begin{split}
\int_{-\infty}^\infty d \varphi \, e^{-\frac{S}{\hbar}}
& = \sqrt{2 \pi} \, \biggl[ \, 1-\frac{3}{4} \, \lambda +\frac{105}{32} \, \lambda^2
-\frac{3465}{128} \, \lambda^3 +\frac{675675}{2048} \, \lambda^4
+\mathcal{O} (\lambda^5) \, \biggr] \,, \\
\int_{-\infty}^\infty d \varphi \, \varphi^2 \, e^{-\frac{S}{\hbar}}
& = \sqrt{2 \pi} \, \biggl[ \, 1-\frac{15}{4} \, \lambda +\frac{945}{32} \, \lambda^2
-\frac{45045}{128} \, \lambda^3 +\frac{11486475}{2048} \, \lambda^4
+\mathcal{O} (\lambda^5) \, \biggr] \,, \\
\int_{-\infty}^\infty d \varphi \, \varphi^4 \, e^{-\frac{S}{\hbar}}
& = \sqrt{2 \pi} \, \biggl[ \, 3 -\frac{105}{4} \, \lambda +\frac{10395}{32} \, \lambda^2
-\frac{675675}{128} \, \lambda^3 +\frac{218243025}{2048} \, \lambda^4
+\mathcal{O} (\lambda^5) \, \biggr] \,,
\end{split}
\end{equation}
the perturbative expansions of
$\langle \, \varphi^2 \, \rangle$ and $\langle \, \varphi^4 \, \rangle$ in the path integral formalism are given by
\begin{equation}
\begin{split}
\langle \, \varphi^2 \, \rangle
& = 1-3 \lambda+24  \lambda^2-297  \lambda^3+4896  \lambda^4+\mathcal{O}\left( \lambda^5\right)\,, \\
\langle \, \varphi^4 \, \rangle
& = 3-24 \lambda+297 \lambda^2-4896 \lambda^3+100278 \lambda^4+\mathcal{O}\left(\lambda^5\right) \,.
\end{split}
\end{equation}

\begin{figure}[h]
\centering\includegraphics[width=12cm]{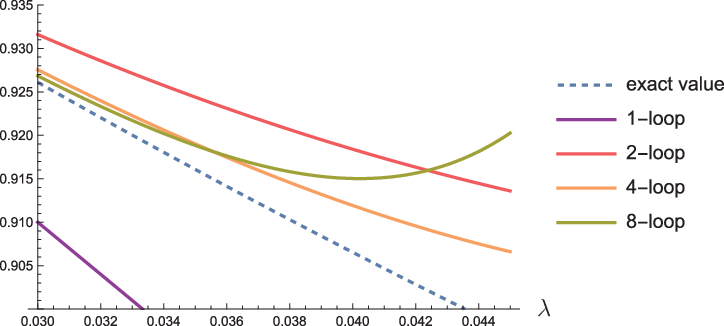}
\caption{The plots of the exact value of the two-point function $\langle\,\varphi^{2}\,\rangle$
and its loop approximations.}
\label{figure-Euclidean-perturbative}
\end{figure}

Let us next calculate the perturbative expansions of correlation functions
in terms of quantum $A_{\infty}$ algebras. In the case of $\varphi^4$ theory in zero dimensions,
$\bm{f}$ is given by
\begin{equation}
\bm{f} = \frac{1}{{\bf I} +{\bm h} \, {\bm m}_3 -\hbar \, {\bm h} \, {\bf U}} \,.
\end{equation}
In perturbation theory, $\bm{f}$ is defined by
\begin{equation}
\bm{f} = \frac{1}{{\bf I} +{\bm h} \, {\bm m}_3 -\hbar \, {\bm h} \, {\bf U}}
= {\bf I} +\sum_{n=1}^\infty \, (-1)^n \,
( \, {\bm h} \, {\bm m}_3 -\hbar \, {\bm h} \, {\bf U} \, )^n \,.
\end{equation}
Note that the coupling constant $\lambda$ is contained in ${\bm h} \, {\bm m}_3$.
We expand $\bm{f}$ as
\begin{equation}
\bm{f} = \frac{1}{{\bf I} -\hbar \, {\bm h} \, {\bf U}}
+\sum_{n=1}^\infty \, (-1)^n \, \frac{1}{{\bf I} -\hbar \, {\bm h} \, {\bf U}} \,
\biggl( \, {\bm h} \, {\bm m}_3 \,
\frac{1}{{\bf I} -\hbar \, {\bm h} \, {\bf U}} \, \biggr)^n \,,
\end{equation}
where
\begin{equation}
\frac{1}{{\bf I} -\hbar \, {\bm h} \, {\bf U}}
= {\bf I} +\sum_{n=1}^\infty \,
( \, \hbar \, {\bm h} \, {\bf U} \, )^n \,.
\end{equation}
When we evaluate the $\mathcal{O} (\lambda^m)$ term of $\bm{f}_{n0}$,
we can replace the inverse of ${\bf I} -\hbar \, {\bm h} \, {\bf U}$ with
\begin{equation}
{\bf I} +\sum_{k=1}^x \,
( \, \hbar \, {\bm h} \, {\bf U} \, )^k \,,
\end{equation}
where $x$ is determined by $m$ and $n$.\footnote{
When $n$ is even, $x$ for $\varphi^4$ theory is given by $x = (2m+n)/2 \,$.
}
We calculate $\bm{f}_{20}$ and $\bm{f}_{40}$ to find
\begin{equation}
\begin{split}
\bm{f}_{20}
& = 1-3 \lambda+24  \lambda^2-297  \lambda^3+4896  \lambda^4+\mathcal{O}\left( \lambda^5\right) \,, \\
\bm{f}_{40}
& = 3-24 \lambda+297 \lambda^2-4896 \lambda^3+100278 \lambda^4+\mathcal{O}\left(\lambda^5\right)\,.
\end{split}
\end{equation}
These correctly produce the perturbative expansions
of $\langle \, \varphi^2 \, \rangle$ and $\langle \, \varphi^4 \, \rangle$.
We show the plots of the exact value of $\langle \, \varphi^2 \, \rangle$
and its perturbative expansions
in figure~\ref{figure-Euclidean-perturbative}.
The perturbative expansion does not converge,
and we see from figure~\ref{figure-Euclidean-perturbative}
that the 8-loop approximation is worse than the 4-loop approximation
when $\lambda \sim 0.04$.
Numerical values of loop approximations
for the two-point function $\langle\,\varphi^{2}\,\rangle$
when $\lambda = 0.04$ are presented up to 8 loops
in table~\ref{table-Euclidean-higher-perturbative}.
\begin{table}[h]
\begin{center}
\begin{tabular}{|c|l|l|}
\hline
$\langle \varphi^2 \rangle$ & \qquad \quad $\lambda=0.04$ & \qquad \qquad \quad $\lambda=0.2$  \\ \hline
exact & $0.906536724355833769$ & \phantom{000}$\phantom{-}0.724059020240824977$  \\ \hline\hline
the number of loops  & \qquad \quad $\lambda=0.04$ & \qquad \qquad \quad $\lambda=0.2$\\ \hline
$1$ & $0.88$ & \phantom{000}$\phantom{-}0.4$\\ \hline
$2$ & $0.9184$ & \phantom{000}$\phantom{-}1.36$\\ \hline
$3$ & $0.899392$ & \phantom{000}$-1.016$\\ \hline
$4$ & $0.91192576$ & \phantom{000}$\phantom{-}6.8176$\\ \hline
$5$ & $0.9016572928$ & \phantom{00}$-25.27136$\\ \hline
$6$ & $0.911693737984$ & \phantom{0}$\phantom{-}131.548096$\\ \hline
$7$ & $0.90030138621952$ & \phantom{0}$-758.4793856$\\ \hline
$8$ & $0.9150305504198656$ & $\phantom{-}4995.10038016$\\ \hline
\end{tabular}
\end{center}
\caption{Evaluation of the two-point function $\langle \, \varphi^2 \, \rangle$
in perturbation theory.}
\label{table-Euclidean-higher-perturbative}
\end{table}

As we mentioned in the introduction,
it is possible that the inverse of
${\bf I} +{\bm h} \, {\bm m}_3 -\hbar \, {\bm h} \, {\bf U}$
exists nonperturbatively for finite values of $\lambda$.
When $m=1$ and $\hbar = 1$, the matrix form of
${\bf I} +{\bm h} \, {\bm m}_3 -\hbar \, {\bm h} \, {\bf U}$ is
\begin{equation}
{\bf I} +{\bm h} \, {\bm m}_3 -\hbar \, {\bm h} \, {\bf U} = \left(
\begin{array}{ccccccc}
1 & 0 & 0 & 0 & 0 & 0 & \ldots \\
0 & 1 & 0 & \lambda & 0 & 0 & \ldots \\
-1 & 0 & 1 & 0 & \lambda & 0 & \ldots \\
0 & -2 & 0 & 1 & 0 & \lambda & \ldots \\
0 & 0 & -3 & 0 & 1 & 0 & \ldots \\
0 & 0 & 0 & -4 & 0 & 1 & \ldots \\
\vdots & \vdots & \vdots & \vdots & \vdots & \vdots & \ddots
\end{array}
\right) \,.
\end{equation}
Let us evaluate the inverse of ${\bf I} +{\bm h} \, {\bm m}_3 -\hbar \, {\bm h} \, {\bf U}$
by truncating it to an $N$ by $N$ matrix.
For $N = 25$, $\bm{f}_{20}$ and $\bm{f}_{40}$ as functions of $\lambda$ are given by
\begin{align}
{\bm f}_{20}
& = \frac{1 + 140 \lambda + 6660 \lambda^2 + 129360 \lambda^3 + 957075 \lambda^4 + 1853460 \lambda^5}
{1 + 143 \lambda + 7065 \lambda^2 + 147420 \lambda^3 + 1267350 \lambda^4 + 3615885 \lambda^5 + 1514205 \lambda^6} \,,
\label{f_20-Euclidean} \\
{\bm f}_{40}
& = \frac{3 + 405 \lambda + 18060 \lambda^2 + 310275 \lambda^3 + 1762425 \lambda^4 + 1514205 \lambda^5}
{1 + 143 \lambda + 7065 \lambda^2 + 147420 \lambda^3 + 1267350 \lambda^4 + 3615885 \lambda^5 + 
 1514205 \lambda^6}  \,.
\label{f_40-Euclidean}
\end{align}
We can estimate from table~\ref{table-Euclidean-higher-perturbative}
that the precision achieved by the perturbative expansion
when $\lambda = 0.04$ is about $0.5 \%$.
With this in mind, let us now substitute $\lambda = 0.04$ in~\eqref{f_20-Euclidean} and~\eqref{f_40-Euclidean},
and compare them with the corresponding exact values of the correlation functions.\footnote{
We performed the integrals analytically using {\it Mathematica}
to obtain expressions in terms of the modified Bessel functions,
and then we presented their numerical values in~\eqref{two-point-exact} and~\eqref{four-point-exact}.
In the rest of this section, we similarly use {\it Mathematica}
for performing integrals analytically and for subsequent numerical evaluations.
} We find
\begin{align}
\langle \, \varphi^2 \, \rangle & \simeq 0.90653672 \,,
\label{two-point-exact} \\
{\bm f}_{20} & \simeq 0.90653666
\end{align}
and
\begin{align}
\langle \, \varphi^4 \, \rangle & \simeq 2.3365819 \,,
\label{four-point-exact} \\
{\bm f}_{40} & \simeq 2.3365834 \,.
\end{align}
The agreement of ${\bm f}_{n0}$ with $\langle \, \varphi^n \, \rangle$
is much better than that of the perturbative expansion.
We do not believe that this is accidental, and we consider this result
as evidence that the formula for correlation functions
based on quantum $A_\infty$ algebras contains information beyond perturbation theory.
\begin{table}[t]
\begin{center}
\begin{tabular}{|c|c|c|c|c|}
\hline
$\langle \, \varphi^2 \, \rangle$ & $\lambda=0.04$ & $\lambda=0.2$ & $\lambda=1.5$& $\lambda=3$ \\ \hline
exact  & $0.9065367244$ & $0.7240590202$ & $0.4066915207$& $0.3130156270$ \\ \hline\hline
$N$  & $\lambda=0.04$ & $\lambda=0.2$ &  $\lambda=1.5$& $\lambda=3$ \\ \hline
$10$ & $0.9059745348$ & $0.7024793388$ & $0.2685512367$ & $0.1574468085$ \\ \hline
$25$ & $0.9065366639$ & $0.7237546945$ & $0.3751623774$ & $0.2543859219$ \\ \hline
$50$  & $0.9065367244$ & $0.7240552164$ & $0.4002397294$& $0.2932452874$ \\ \hline
$100$  & $0.9065367244$ & $0.7240590258$ & $0.4072861268$& $0.3167705780$ \\ \hline
\end{tabular}
\end{center}
\caption{Evaluation of ${\bm f}_{20}$ by truncating
${\bf I} +{\bm h} \, {\bm m}_3 -\hbar \, {\bm h} \, {\bf U}$ to an $N$ by $N$ matrix
for various values of $N$
and comparison with the two-point function $\langle \varphi^2 \rangle$.}
\label{table-Euclidean-nonperturbative1}
\end{table}

Let us explore larger values of $\lambda$.
The data for the perturbative expansion of the two-point function
when $\lambda = 0.2$ are also presented in table~\ref{table-Euclidean-higher-perturbative},
and we see that the perturbation theory completely breaks down.
For $\lambda = 0.2$, the evaluation of ${\bm f}_{20}$ with $N=100$
and the exact value of $\langle\, \varphi^2\, \rangle$ are
\begin{align}
\langle\, \varphi^2 \, \rangle & \simeq 0.7240590202 \,, \\
{\bm f}_{20} & \simeq 0.7240590258 \,.
\end{align}
Even when the perturbative expansion breaks down,
we see that the formula for correlation functions based on quantum $A_\infty$ algebras
still works!
While requiring a precision of about $1 \%$, we can increase the coupling constant
up to $\lambda \simeq 3$ when $N = 100 \,$.
Further data are presented in table~\ref{table-Euclidean-nonperturbative1}
and table~\ref{table-Euclidean-nonperturbative2}.
\begin{table}[t]
\begin{center}
\begin{tabular}{|c|c|c|c|c|}
\hline
$\langle \, \varphi^4 \, \rangle$ & $\lambda=0.04$ & $\lambda=0.2$ & $\lambda=1.5$& $\lambda=3$ \\ \hline
exact & $2.336581891$ & $1.379704899$ & $0.3955389862$& $0.2289947910$ \\ \hline\hline
$N$  & $\lambda=0.04$ & $\lambda=0.2$ & $\lambda=1.5$& $\lambda=3$  \\ \hline
$10$  & $2.350636631$ & $1.487603306$ & $0.4876325088$& $0.2808510638$ \\ \hline
$25$  & $2.336583402$ & $1.381226527$ & $0.4165584151$& $0.2485380260$ \\ \hline
$50$  & $2.336581891$ & $1.379723918$ & $0.3998401804$& $0.2355849042$ \\ \hline
$100$  & $2.336581891$ & $1.379704871$ & $0.3951425821$& $0.2277431407$ \\ \hline
\end{tabular}
\end{center}
\caption{Evaluation of ${\bm f}_{40}$ by truncating
${\bf I} +{\bm h} \, {\bm m}_3 -\hbar \, {\bm h} \, {\bf U}$ to an $N$ by $N$ matrix
for various values of $N$
and comparison with the four-point function $\langle \varphi^4 \rangle$.}
\label{table-Euclidean-nonperturbative2}
\end{table}

\subsection{The Lorentzian case}\label{Lorentzian-quartic}

Let us next consider the Lorentzian case.
As in the Euclidean case, we set $g=0$ and consider $\varphi^4$ theory.
The formula for correlation functions in the Lorentzian case
is simply obtained by replacing $-\hbar \, {\bm h} \, {\bf U}$
in the Euclidean case with $i \hbar \, {\bm h} \, {\bf U}$:
\begin{equation}
\langle \, \varphi^n \, \rangle
= \bm{f}_{n0} \,,
\end{equation}
where
\begin{equation}
\bm{f}
= \frac{1}{{\bf I} +{\bm h} \, {\bm m}_3 +i \hbar \, {\bm h} \, {\bf U}} \,.
\end{equation}
The definitions of ${\bm h} \, {\bm m}_3$ and ${\bm h} \, {\bf U}$ are the same
as those in the preceding subsection.
The formula reproduces perturbation theory,
although we do not present it here.
Again, it is possible that the inverse of the operator
${\bf I} +{\bm h} \, {\bm m}_3 +i \hbar \, {\bm h} \, {\bf U}$
exists nonperturbatively for finite values of $\lambda$.
When $m=1$ and $\hbar = 1$, the matrix form of
${\bf I} +{\bm h} \, {\bm m}_3 +i \hbar \, {\bm h} \, {\bf U}$ is
\begin{equation}
{\bf I} +{\bm h} \, {\bm m}_3 +i \hbar \, {\bm h} \, {\bf U} = \left(
\begin{array}{ccccccc}
1 & 0 & 0 & 0 & 0 & 0 & \ldots \\
0 & 1 & 0 & \lambda & 0 & 0 & \ldots \\
i & 0 & 1 & 0 & \lambda & 0 & \ldots \\
0 & 2i & 0 & 1 & 0 & \lambda & \ldots \\
0 & 0 & 3i & 0 & 1 & 0 & \ldots \\
0 & 0 & 0 & 4i & 0 & 1 & \ldots \\
\vdots & \vdots & \vdots & \vdots & \vdots & \vdots & \ddots
\end{array}
\right) \,.
\end{equation}
Let us evaluate the inverse of ${\bf I} +{\bm h} \, {\bm m}_3 +i \hbar \, {\bm h} \, {\bf U}$
by truncating it to an $N$ by $N$ matrix.
For $N = 25$, $\bm{f}_{20}$ and $\bm{f}_{40}$ as functions of $\lambda$ are given by
\begin{align}
{\bm f}_{20} & = \frac
{-i - 140 \lambda + 6660 i \lambda^2 + 129360 \lambda^3 - 957075 i \lambda^4 - 1853460 \lambda^5}
{1 - 143 i  \lambda  - 7065  \lambda^2 + 147420 i  \lambda^3 + 1267350  \lambda^4 
- 3615885 i  \lambda^5 - 1514205  \lambda^6} \,,
\label{f_20-Lorentzian} \\
{\bm f}_{40} & = \frac
{-3 + 405 i \lambda + 18060 \lambda^2 - 310275 i \lambda^3 - 1762425 \lambda^4 + 1514205 i \lambda^5}
{1 - 143 i  \lambda  - 7065  \lambda^2 + 147420 i  \lambda^3 + 1267350  \lambda^4 
- 3615885 i  \lambda^5 - 1514205  \lambda^6} \,.
\label{f_40-Lorentzian}
\end{align}
As in the Euclidean case,
let us substitute $\lambda = 0.04$ in~\eqref{f_20-Lorentzian} and~\eqref{f_40-Lorentzian},
and compare them with the corresponding exact values of the correlation functions. We find
\begin{align}
\langle \, \varphi^2 \, \rangle & \simeq 0.1065670 - 0.969384893 i \,, \\
{\bm f}_{20} & \simeq 0.1065659 - 0.969384873 i
\end{align}
and
\begin{align}
\langle \, \varphi^4 \, \rangle & \simeq -2.6641739 - 0.7653777 i \,, \\
{\bm f}_{40} & \simeq -2.6641477 - 0.7653782 i \,.
\end{align}
As in the Euclidean case,
we find impressive agreement.
We consider that the formula for correlation functions
based on quantum $A_\infty$ algebras contains information beyond perturbation theory
in the Lorentzian case as well.

\begin{table}
\begin{center}
\begin{tabular}{|c|c|c|}
\hline 
$\langle \, \varphi^2 \, \rangle$ &$\lambda=0.04$ & $\lambda=0.5$ \\ \hline
exact & $0.106567 - 0.969385 i$ & $0.280132 - 0.576152 i$ \\ \hline\hline
$N$ &  $\lambda=0.04$ & $\lambda=0.5$ \\ \hline
$10$ & $0.105205 - 0.969450 i$ & $0.444071 - 0.548360 i$  \\ \hline 
$25$ & $0.106566 - 0.969385 i$ & $0.274637 - 0.597967 i$   \\ \hline 
$50$ & $0.106567 - 0.969385 i$ & $0.277626 - 0.574320 i$   \\ \hline 
$100$ & $0.106567 - 0.969385 i$ & $0.279980 - 0.576085 i$   \\ \hline
\end{tabular}
\end{center}
\caption{Evaluation of ${\bm f}_{20}$ by truncating
${\bf I} +{\bm h} \, {\bm m}_3 +i \hbar \, {\bm h} \, {\bf U}$ to an $N$ by $N$ matrix
for various values of $N$
and comparison with the two-point function $\langle \varphi^2 \rangle$.}
\label{table-Lorentzian-nonperturbative1}
\end{table}

As a larger value of $\lambda$, let us take $\lambda = 0.5$.
For $N = 100$, we find
\begin{align}
\langle \, \varphi^2 \, \rangle & \simeq 0.28013 - 0.57615 i \,, \\
{\bm f}_{20} & \simeq 0.27998 - 0.57609 i
\end{align}
and
\begin{align}
\langle \, \varphi^4 \, \rangle & \simeq -0.56026 - 0.84770 i \,, \\
{\bm f}_{40} & \simeq -0.55996 - 0.84783 i \,.
\end{align}
The perturbative expansion breaks down for this value of the coupling constant,
but we again see that the formula for correlation functions based on quantum $A_\infty$ algebras
is working.
Further data are presented in table~\ref{table-Lorentzian-nonperturbative1}
and table~\ref{table-Lorentzian-nonperturbative2}.

The convergence of the path integral in the Lorentzian case is more subtle
than that in the Euclidean case.
On the other hand, the calculations in the formula for correlation functions
based on quantum $A_\infty$ algebras in the Lorentzian case
do not seem to be so different from those in the Euclidean case.
We hope that our approach will provide a new perspective
on the sign problem in the path integral.

\begin{table}
\begin{center}
\begin{tabular}{|c|c|c|}
\hline 
$\langle \, \varphi^4 \, \rangle$ &$\lambda=0.04$ & $\lambda=0.5$ \\ \hline
exact & $-2.66417 - 0.76538 i$ & $-0.560264 - 0.847696 i$ \\ \hline\hline
$N$  &$\lambda=0.04$ & $\lambda=0.5$ \\ \hline
$10$ & $-2.63013 - 0.76374 i$ & $-0.888141 - 0.903280 i$  \\ \hline 
$25$ & $-2.66415 - 0.76538 i$ & $-0.549273 - 0.804066 i$   \\ \hline 
$50$ & $-2.66417 - 0.76538 i$ & $-0.555251 - 0.851360 i$   \\ \hline 
$100$ & $-2.66417 - 0.76538 i$ & $-0.559961 - 0.847830 i$   \\ \hline
\end{tabular}
\end{center}
\caption{Evaluation of ${\bm f}_{40}$ by truncating
${\bf I} +{\bm h} \, {\bm m}_3 +i \hbar \, {\bm h} \, {\bf U}$ to an $N$ by $N$ matrix
for various values of $N$
and comparison with the four-point function $\langle \varphi^4 \rangle$.}
\label{table-Lorentzian-nonperturbative2}
\end{table}

\section{New formula for correlation functions}
\setcounter{equation}{0}

We have presented evidence that our formula for correlation functions
based on quantum $A_\infty$ algebras
contains nonperturbative information.
Our formula, however, requires us to choose a free part of the action.
This is not satisfactory for a nonperturbative definition of correlation functions.
This also raises the question of whether our formula is background independent.
Namely, the question is whether our formula gives the same answer
when there are multiple solutions to the equations of motion
and we choose different free parts.
In this section we give an answer to this question.
We first consider perturbation theory around a nontrivial solution
in subsection~\ref{nontrivial-vacuum-section}.
We then transform the formula to a new form
which does not involve the division of the action into the free part and the interaction part
in~subsection~\ref{transforming-section}.
The new form of the formula requires us to choose a solution to the equations of motion
which does not have to be real.
We state our claim in subsection~\ref{interpretation-section}
that the formula gives correlation functions
on the Lefschetz thimble associated with the solution we chose.
In this section, we consider the Lorentzian case,
but it is straightforward to modify the discussion to the Euclidean case.
   
\subsection{Perturbation theory around a nontrivial solution}
\label{nontrivial-vacuum-section}

\begin{figure}
\begin{center}
\includegraphics[width=7cm]{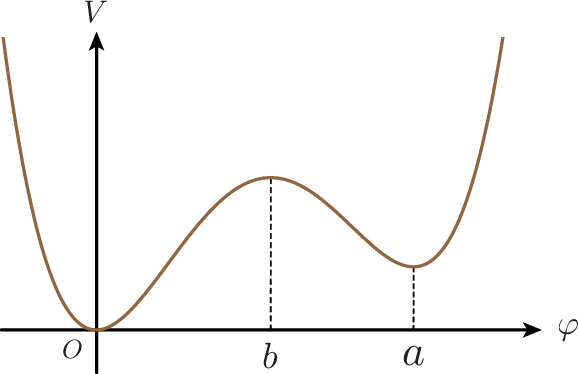}
\end{center}
\caption{The shape of the potential for~\eqref{two-minimum-example}.}
\label{figure-potential}
\end{figure}

To discuss perturbation theory around a nontrivial solution,
let us consider an action given by
\begin{equation}
S = {}-\frac{1}{2} \, m^2 \, \varphi^2
+\frac{(a+b) \, m^2}{3ab} \, \varphi^3 -\frac{m^2}{4ab} \, \varphi^4 \,,
\label{two-minimum-example}
\end{equation}
where the constants $m$, $a$ and $b$ are real and positive with $b < a$.
The shape of the potential is presented in figure~\ref{figure-potential}.\footnote{
The potential $V$ is related to the action $S$ as $V = {}-S$ in the Lorentzian case
and as $V = S$ in the Euclidean case.
}
The equation of motion is given by
\begin{equation}
\frac{m^2}{ab} \, \varphi \, ( \varphi -b ) \, ( \varphi -a ) = 0 \,,
\end{equation}
and the solutions are
\begin{equation}
\varphi = 0 \,, b \,, a \,. 
\end{equation}
We know the formula for perturbation theory around the solution $\varphi = 0$.
It corresponds to the case where
\begin{equation}
Q \, c = m^{2} \, d \,, \quad
m_2 \, ( \, c \otimes c \, ) = {}-\frac{(a+b) \, m^2}{ab} \, d \,, \quad
m_3 \, ( \, c \otimes c \otimes c \, ) = \frac{m^2}{ab} \,  d \,, \quad
h \, d = \frac{1}{m^2} \, c \,,
\end{equation}
and the correlation functions are given by
\begin{equation}
\langle \, \Phi^{\otimes n} \, \rangle
= \pi_n \, \frac{1}{{\bf I} +{\bm h} \, {\bm m} +i \hbar \, {\bm h} \, {\bf U}} \, {\bf 1} \,.
\end{equation}

Let us consider perturbation theory around the nontrivial solution $\varphi = a$.
In quantum field theory, we know what to do.
We expand $\varphi$ as
\begin{equation}
\varphi = a +\widetilde{\varphi} \,,
\end{equation}
and the action in terms of $\widetilde{\varphi}$ is
\begin{equation}
S = {}-\frac{a^2 \, (2b-a) \, m^2}{12b}
-\frac{(a-b) \, m^2}{2b} \, \widetilde{\varphi}^{\, 2}
-\frac{(2a-b) \, m^2}{3ab} \, \widetilde{\varphi}^{\, 3}
-\frac{m^2}{4ab} \, \widetilde{\varphi}^{\, 4} \,.
\label{varphi-tilde-action}
\end{equation}
We then calculate $\langle \, \widetilde{\varphi}^{\, n} \, \rangle$,
and $\langle \, \varphi^n \, \rangle$ is given by
\begin{equation}
\langle \, \varphi^n \, \rangle
= \langle \, ( \, a+\widetilde{\varphi} \, )^n \, \rangle
= \sum_{m=0}^n \, \frac{n!}{m! \, (n-m)!} \,
a^{n-m} \, \langle \, \widetilde{\varphi}^{\, m} \, \rangle \,.
\end{equation}

Let us describe this procedure in terms of $A_\infty$ algebras.
We first need to represent the equations of motion
in the language of $A_\infty$ algebras.
The equations of motion are usually written as
\begin{equation}
\pi_1 \, {\bf M} \, \frac{1}{1 -\Phi} = 0 \,,
\end{equation}
where
\begin{equation}
\frac{1}{1 -\Phi} = \sum_{n=0}^\infty \Phi^{\otimes n}
= {\bf 1} +\Phi +\Phi \otimes \Phi +\Phi \otimes \Phi \otimes \Phi + \ldots \,.
\end{equation}
An element of this form is often called a group-like element.
Note that this satisfies
\begin{equation}
\triangle \, \frac{1}{1 -\Phi} = \frac{1}{1 -\Phi} \otimes' \frac{1}{1 -\Phi} \,.
\end{equation}
This can also be written in a form which is more convenient for us
using the coderivation ${\bf \Phi}$ associated with $\Phi$ in~\eqref{Phi-coderivation}.
Since
\begin{equation}
\triangle \, e^{\bf \Phi} \, {\bf 1}
= e^{\bf \Phi} \, {\bf 1} \otimes' e^{\bf \Phi} \, {\bf 1} \,, \qquad
\pi_1 \, e^{\bf \Phi} \, {\bf 1} = \Phi \,,
\end{equation}
we have
\begin{equation}
\frac{1}{1 -\Phi} = e^{\bf \Phi} \, {\bf 1} \,.
\end{equation}
In terms of ${\bf \Phi}$,
the equations of motion are written as
\begin{equation}
\pi_1 \, {\bf M} \, e^{\bf \Phi} \, {\bf 1} = 0 \,.
\end{equation}
In the case of scalar field theories in zero dimensions,
we have
\begin{equation}
{\bf \Phi} \, {\bf 1} = \Phi = \varphi \, c \,.
\end{equation}
For the action~\eqref{two-minimum-example}, we find
\begin{equation}
\begin{split}
\pi_1 \, {\bf Q} \, {\bf \Phi} \, {\bf 1}
& = Q \, \Phi = m^2 \, \varphi \, d \,, \\
\frac{1}{2} \, \pi_1 \, {\bm m}_2 \, {\bf \Phi}^2 \, {\bf 1}
& = \frac{1}{2} \, m_2 \, ( \, \Phi \otimes \mathbb{I} +\mathbb{I} \otimes \Phi \, ) \, \Phi
= g \, \varphi^2 \, d \,, \\
\frac{1}{3!} \, \pi_1 \, {\bm m}_3 \, {\bf \Phi}^3 \, {\bf 1}
& = \frac{1}{3!} \, m_3 \,
( \, \Phi \otimes \mathbb{I} \otimes \mathbb{I}
+\mathbb{I} \otimes \Phi \otimes \mathbb{I}
+\mathbb{I} \otimes \mathbb{I} \otimes \Phi \, ) \,
( \, \Phi \otimes \mathbb{I} +\mathbb{I} \otimes \Phi \, ) \, \Phi \\
& = \lambda \, \varphi^3 \, d \,.
\end{split}
\end{equation}
Therefore, the equation of motion is reproduced as follows:
\begin{equation}
\pi_1 \, {\bf M} \, e^{\bf \Phi} \, {\bf 1}
= ( \, m^2 \, \varphi +g \, \varphi^2 +\lambda \, \varphi^3 \, ) \, d = 0 \,.
\end{equation}

Suppose that we have a nontrivial solution $\Phi_\ast$ to the equations of motion.
We denote the coderivation associated with $\Phi_\ast$ by ${\bf \Phi}_\ast$:
\begin{equation}
\pi_1 \, {\bf M} \, e^{{\bf \Phi}_\ast} \, {\bf 1} = 0 \,.
\label{coderivation-Phi_ast-equation}
\end{equation}
In the case of the theory~\eqref{two-minimum-example}
with the solution $\varphi = a$, we have
\begin{equation}
\Phi_\ast = a \, c \,.
\end{equation}
We expand $\Phi$ as
\begin{equation}
\Phi = \Phi_\ast +\widetilde{\Phi} \,,
\end{equation}
and let us consider
the coderivation $\widetilde{\bf M}$ which describes the action in terms of $\widetilde{\Phi}$.
We take a generic quartic theory as an example for illustration.
The operator $\widetilde{m}_3$ is the same as $m_3$:
\begin{equation}
\widetilde{m}_3 = m_3 \,.
\end{equation}
We therefore have
\begin{equation}
\pi_1 \, \widetilde{\bm m}_3 = \pi_1 \, {\bm m}_3 \,.
\end{equation}
The operator $\widetilde{m}_2$ is given by
\begin{equation}
\widetilde{m}_2 = m_2 +m_3 \,
( \, \Phi_\ast \otimes \mathbb{I} \otimes \mathbb{I}
+\mathbb{I} \otimes \Phi_\ast \otimes \mathbb{I}
+\mathbb{I} \otimes \mathbb{I} \otimes \Phi_\ast \, ) \,.
\end{equation}
We therefore have
\begin{equation}
\pi_1 \, \widetilde{\bm m}_2
= \pi_1 \, {\bm m}_2 +\pi_1 \, {\bm m}_3 \, {\bf \Phi}_\ast \,.
\end{equation}
The operator $\widetilde{Q}$ is given by
\begin{equation}
\widetilde{Q} = Q +m_2 \, ( \, \Phi_\ast \otimes \mathbb{I} +\mathbb{I} \otimes \Phi_\ast \, )
+m_3 \,
( \, \Phi_\ast \otimes \Phi_\ast \otimes \mathbb{I}
+\Phi_\ast \otimes \mathbb{I} \otimes \Phi_\ast
+\mathbb{I} \otimes \Phi_\ast \otimes \Phi_\ast \, ) \,.
\end{equation}
Since
\begin{equation}
\begin{split}
& \Phi_\ast \otimes \Phi_\ast \otimes \mathbb{I}
+\Phi_\ast \otimes \mathbb{I} \otimes \Phi_\ast
+\mathbb{I} \otimes \Phi_\ast \otimes \Phi_\ast \\
& = \frac{1}{2} \,
( \, \Phi_\ast \otimes \mathbb{I} \otimes \mathbb{I}
+\mathbb{I} \otimes \Phi_\ast \otimes \mathbb{I}
+\mathbb{I} \otimes \mathbb{I} \otimes \Phi_\ast \, ) \,
( \, \Phi_\ast \otimes \mathbb{I} +\mathbb{I} \otimes \Phi_\ast \, ) \,,
\end{split}
\end{equation}
we have
\begin{equation}
\pi_1 \, \widetilde{\bf Q}
= \pi_1 \, {\bf Q}
+\pi_1 \, {\bm m}_2 \, {\bf \Phi}_\ast
+\frac{1}{2} \, \pi_1 \, {\bm m}_3 \, {\bf \Phi}_\ast^2 \,.
\end{equation}
Let us confirm this for the theory~\eqref{two-minimum-example} with the solution $\varphi = a$.
The action~\eqref{varphi-tilde-action} is described by
\begin{equation}
\widetilde{Q} \, c = \frac{(a-b) \, m^2}{b} \, d \,, \quad
\widetilde{m}_2 \, ( \, c \otimes c \, ) = \frac{(2a-b) \, m^2}{ab} \, d \,, \quad
\widetilde{m}_3 \, ( \, c \otimes c \otimes c \, ) = \frac{m^2}{ab} \,  d \,.
\end{equation}
The operator $\widetilde{m}_3$ is the same as $m_3$.
For $\widetilde{m}_2$, we have
\begin{equation}
\begin{split}
& m_2 \, ( \, c \otimes c \, )
+m_3 \,
( \, a \, c \otimes \mathbb{I} \otimes \mathbb{I}
+\mathbb{I} \otimes a \, c  \otimes \mathbb{I}
+\mathbb{I} \otimes \mathbb{I} \otimes a \, c  \, ) \,
( \, c \otimes c \, ) \\
& = {}-\frac{(a+b) \, m^2}{ab} \, d +\frac{3 m^2}{b} \, d
= \frac{(2a-b) \, m^2}{ab} \, d \,,
\end{split}
\end{equation}
which reproduces $\widetilde{m}_2 \, ( \, c \otimes c \, ) \,$.
For $\widetilde{Q}$, we have
\begin{equation}
\begin{split}
& Q \, c +m_2 \, ( \, a \, c \otimes \mathbb{I} +\mathbb{I} \otimes a \, c \, ) \, c \\
& +\frac{1}{2} \, m_3 \,
( \, a \, c \otimes \mathbb{I} \otimes \mathbb{I}
+\mathbb{I} \otimes a \, c  \otimes \mathbb{I}
+\mathbb{I} \otimes \mathbb{I} \otimes a \, c  \, ) \,
( \, a \, c \otimes \mathbb{I} +\mathbb{I} \otimes a \, c  \, ) \, c \\
& = m^2 \, d
-\frac{2 \, (a+b) \, m^2}{b} \, d +\frac{3 a \, m^2}{b} \, d
= \frac{(a-b) \, m^2}{b} \, d \,,
\end{split}
\end{equation}
which reproduces $\widetilde{Q} \, c \,$.

As can be seen from this discussion for the quartic theory,
we in general have
\begin{equation}
\pi_1 \, \widetilde{\bf M}
= \pi_1 \, {\bf M} \, e^{{\bf \Phi}_\ast} \,,
\end{equation}
and the coderivation $\widetilde{\bf M}$ is given by
\begin{equation}
\widetilde{\bf M}
= e^{{}-{\bf \Phi}_\ast} \, {\bf M} \, e^{{\bf \Phi}_\ast}
\end{equation}
since $\triangle \, ( \, e^{{}-{\bf \Phi}_\ast} \, {\bf M} \, e^{{\bf \Phi}_\ast} \, )
= ( \, e^{{}-{\bf \Phi}_\ast} \, {\bf M} \, e^{{\bf \Phi}_\ast} \otimes' {\bf I}
+{\bf I} \otimes' e^{{}-{\bf \Phi}_\ast} \, {\bf M} \, e^{{\bf \Phi}_\ast} \, ) \, \triangle$
and $\pi_1 \, e^{{}-{\bf \Phi}_\ast} \, {\bf M} \, e^{{\bf \Phi}_\ast}
= \pi_1 \, {\bf M} \, e^{{\bf \Phi}_\ast}$.
In general, we decompose $\pi_1 \, \widetilde{\bf M}$ as
\begin{equation}
\pi_ 1 \, \widetilde{\bf M}
= \widetilde{Q} \, \pi_1
+\sum_{n=2}^\infty \widetilde{m}_n \, \pi_n \,,
\end{equation}
and we define the coderivation $\widetilde{\bf Q}$
associated with $\widetilde{Q}$
and the coderivation $\widetilde{\bm m}_n$
associated with $\widetilde{m}_n$ for each $n$.
Note that $\pi_ 1 \, \widetilde{\bf M} \, \pi_0$ vanishes,
\begin{equation}
\pi_ 1 \, \widetilde{\bf M} \, \pi_0 = 0 \,,
\end{equation}
as ${\bf \Phi}_\ast$ satisfies~\eqref{coderivation-Phi_ast-equation}.
We thus have
\begin{equation}
\widetilde{\bf M}
= \widetilde{\bf Q} +\widetilde{\bm m}
\end{equation}
with
\begin{equation}
\widetilde{\bm m} = \sum_{n=2}^\infty \widetilde{\bm m}_n \,.
\end{equation}
We then construct $\widetilde{h}$ satisfying
\begin{equation}
\widetilde{Q} \, \widetilde{h} +\widetilde{h} \, \widetilde{Q} = \mathbb{I} \,, \qquad
\widetilde{h}^2 = 0 \,,
\end{equation}
and $\widetilde{\bm h}$ satisfying
\begin{equation}
\widetilde{\bf Q} \, \widetilde{\bm h} +\widetilde{\bm h} \, \widetilde{\bf Q}
= {\bf I} -{\bf P} \,, \qquad
\widetilde{\bm h} \, {\bf P} = 0 \,, \qquad
{\bf P} \, \widetilde{\bm h} = 0 \,, \qquad
\widetilde{\bm h}^2 = 0
\end{equation}
with
\begin{equation}
{\bf P} = \pi_0 \,.
\end{equation}
The correlation function~$\langle \, \widetilde{\Phi}^{\otimes n} \, \rangle$
for the perturbation theory
around the solution $\Phi_\ast$ is given by
\begin{equation}
\langle \, \widetilde{\Phi}^{\otimes n} \, \rangle
= \pi_n \, \frac{1}{{\bf I} +\widetilde{\bm h} \, \widetilde{\bm m}
+i \hbar \, \widetilde{\bm h} \, {\bf U}} \, {\bf 1} \,.
\end{equation}
The correlation function~$\langle \, \Phi^{\otimes n} \, \rangle$
for the perturbation theory
around the solution $\Phi_\ast$ can be written using ${\bf \Phi}_\ast$ as
\begin{equation}
\langle \, \Phi^{\otimes n} \, \rangle
= \pi_n \, e^{{\bf \Phi}_\ast}
\frac{1}{{\bf I} +\widetilde{\bm h} \, \widetilde{\bm m}
+i \hbar \, \widetilde{\bm h} \, {\bf U}} \, {\bf 1} \,.
\label{tilde-formula}
\end{equation}
We write this as
\begin{equation}
\langle \, \Phi^{\otimes n} \, \rangle
= \pi_n \,
\frac{1}{{\bf I}
+( \, e^{{\bf \Phi}_\ast} \, \widetilde{\bm h} \, e^{-{\bf \Phi}_\ast} \, ) \,
( \, e^{{\bf \Phi}_\ast} \, \widetilde{\bm m} \, e^{-{\bf \Phi}_\ast} \, )
+i \hbar \, ( \, e^{{\bf \Phi}_\ast} \, \widetilde{\bm h} \, e^{-{\bf \Phi}_\ast} \, ) \,
( \, e^{{\bf \Phi}_\ast} \, {\bf U} \, e^{-{\bf \Phi}_\ast} \, )} \,
e^{{\bf \Phi}_\ast} \, {\bf 1} \,,
\end{equation}
and we define ${\bf Q}_\ast$, ${\bm m}_\ast$ and ${\bm h}_\ast$ by
\begin{equation}
{\bf Q}_\ast = e^{{\bf \Phi}_\ast} \, \widetilde{\bf Q} \, e^{-{\bf \Phi}_\ast} \,, \qquad
{\bm m}_\ast = e^{{\bf \Phi}_\ast} \, \widetilde{\bm m} \, e^{-{\bf \Phi}_\ast} \,, \qquad
{\bm h}_\ast = e^{{\bf \Phi}_\ast} \, \widetilde{\bm h} \, e^{-{\bf \Phi}_\ast} \,.
\end{equation}
For the operator ${\bf U}$, it follows from
\begin{equation}
[ \, {\bf \Phi}_\ast \,,\, {\bf U} \, ] = 0
\end{equation}
that
\begin{equation}
e^{{\bf \Phi}_\ast} \, {\bf U} \, e^{-{\bf \Phi}_\ast} = {\bf U} \,.
\end{equation}
Here and in what follows, $[ \, A \,,\, B \, ]$
is the graded commutator of $A$ and $B$ with respect to degree.
The formula for $\langle \, \Phi^{\otimes n} \, \rangle$ is then
\begin{equation}
\langle \, \Phi^{\otimes n} \, \rangle
= \pi_n \, \frac{1}{{\bf I} +{\bm h}_\ast \, {\bm m}_\ast
+i \hbar \, {\bm h}_\ast \, {\bf U}} \, {\bf P} \, {\bf 1}
\label{ast-formula}
\end{equation}
with
\begin{equation}
{\bf P} = e^{{\bf \Phi}_\ast} \pi_0 \,.
\end{equation}
Since
\begin{equation}
\widetilde{\bf M}
= \widetilde{\bf Q} +\widetilde{\bm m}
= e^{{}-{\bf \Phi}_\ast} \, {\bf M} \, e^{{\bf \Phi}_\ast} \,,
\end{equation}
we find
\begin{equation}
{\bf Q}_\ast +{\bm m}_\ast = {\bf M} \,.
\end{equation}
Namely, the sum of ${\bf Q}_\ast$ and ${\bm m}_\ast$ is the same as the sum
of ${\bf Q}$ and ${\bm m}$, but ${\bf Q}_\ast$ is different from ${\bf Q}$:
\begin{equation}
{\bf Q}_\ast \ne {\bf Q} \,.
\end{equation}
Given the action described by ${\bf M}$,
we are making different choices for the free part.
If we take ${\bf Q}$ to be the free part,
the correlation functions are given by
\begin{equation}
\langle \, \Phi^{\otimes n} \, \rangle
= \pi_n \, \frac{1}{{\bf I} +{\bm h} \, {\bm m} +i \hbar \, {\bm h} \, {\bf U}} \, {\bf 1} \,.
\end{equation}
If we take ${\bf Q}_\ast$ to be the free part,
the correlation functions are given by
\begin{equation}
\langle \, \Phi^{\otimes n} \, \rangle
= \pi_n \, \frac{1}{{\bf I} +{\bm h}_\ast \, {\bm m}_\ast
+i \hbar \, {\bm h}_\ast \, {\bf U}} \, {\bf P} \, {\bf 1} \,.
\end{equation}
An important lesson we have learned is that in general ${\bf P}$ is different from $\pi_0$.
Note that
\begin{equation}
{\bf P}^2 = {\bf P}
\end{equation}
because
\begin{equation}
e^{{\bf \Phi}_\ast} \, \pi_0 \, e^{{\bf \Phi}_\ast} \pi_0
= e^{{\bf \Phi}_\ast} \, \pi_0 \,,
\end{equation}
where we used $\pi_0 \, {\bf \Phi}_\ast = 0 \,$.
In fact, it is sometimes convenient to think of ${\bf P}$ as
\begin{equation}
{\bf P} = e^{{\bf \Phi}_\ast} \, \pi_0 \, e^{-{\bf \Phi}_\ast} \,,
\end{equation}
although the factor of $e^{-{\bf \Phi}_\ast}$ reduces to the identity operator
because of the action by $\pi_0$.
For example, it follows from
\begin{equation}
{\bf Q} \, {\bm h} +{\bm h} \, {\bf Q} = {\bf I} -{\bf P} \,, \qquad
{\bm h} \, {\bf P} = 0 \,, \qquad
{\bf P} \, {\bm h} = 0 \,, \qquad
{\bm h}^2 = 0
\end{equation}
with ${\bf P} = \pi_0$ that
\begin{equation}
{\bf Q}_\ast \, {\bm h}_\ast +{\bm h}_\ast \, {\bf Q}_\ast = {\bf I} -{\bf P} \,, \qquad
{\bm h}_\ast \, {\bf P} = 0 \,, \qquad
{\bf P} \, {\bm h}_\ast = 0 \,, \qquad
{\bm h}_\ast^2 = 0
\end{equation}
with ${\bf P} = e^{{\bf \Phi}_\ast} \, \pi_0 \, e^{-{\bf \Phi}_\ast}$.

We also define the operator ${\bf P}$ for a general element of $\mathcal{H}_1$
which is not necessarily a solution to the equation of motion.
Then the equations of motion can be written in terms of ${\bf P}$ as
\begin{equation}
{\bf M} \, {\bf P} = 0
\end{equation}
with
\begin{equation}
{\bf P} = e^{\bf \Phi} \, \pi_0 \,,
\end{equation}
where ${\bf \Phi}$ is the coderivation associated with ${\Phi}$.

Practically, the formula of the form~\eqref{tilde-formula} is convenient.
Conceptually, however, the formula of the form~\eqref{ast-formula} is important,
and this form will lead us to a formula which does not involve
the division of the action into the free part and the interaction part
in the next subsection.

\subsection{Transforming the formula}\label{transforming-section}

Let us recapitulate the result of the preceding subsection
in terms of slightly modified notation.
We denote the coderivation which describes the action by ${\bf M}$.
We then choose its free part and a solution to the free equations of motion.
Since it is no longer important whether or not the solution is trivial,
we denote the free part of ${\bf M}$ by ${\bf Q}$
without distinguishing between ${\bf Q}$ and ${\bf Q}_\ast$.
We denote the coderivation which describes the interactions by ${\bm m}$,
and ${\bf M}$ is the sum of ${\bf Q}$ and ${\bm m}$:
\begin{equation}
{\bf M} = {\bf Q} +{\bm m} \,.
\end{equation}
We will later consider the case where the solution solves the free equation of motion
but does not solve the full equations of motion,
so we denote the solution to the free part of the equations of motion by $\Phi_\ast^{(0)}$
instead of $\Phi_\ast$ in the preceding subsection.
In terms of the coderivation ${\bf \Phi}_\ast^{(0)}$ associated with $\Phi_\ast^{(0)}$,
we have
\begin{equation}
{\bf Q} \, {\bf P}^{(0)} = 0
\label{Phi^(0)-coderivation}
\end{equation}
with
\begin{equation}
{\bf P}^{(0)} = e^{{\bf \Phi}_\ast^{(0)}} \pi_0 \,.
\label{free-P}
\end{equation}
We then use ${\bm h}$ which satisfies
\begin{equation}
{\bf Q} \, {\bm h} +{\bm h} \, {\bf Q} = {\bf I} -{\bf P}^{(0)} \,, \qquad
{\bm h} \, {\bf P}^{(0)} = 0 \,, \qquad
{\bf P}^{(0)} \, {\bm h} = 0 \,, \qquad
{\bm h}^2 = 0
\end{equation}
to describe the correlation functions as follows:
\begin{equation}
\langle \, \Phi^{\otimes n} \, \rangle
= \pi_n \, \frac{1}{{\bf I} +{\bm h} \, {\bm m}
+i \hbar \, {\bm h} \, {\bf U}} \, {\bf P}^{(0)} \, {\bf 1} \,.
\end{equation}

Let us transform the formula into a form that does not involve
the division of the action into the free part and the interaction part.
Since
\begin{equation}
{\bf I} +{\bm h} \, {\bm m} +i \hbar \, {\bm h} \, {\bf U}
= ( \, {\bf I} +{\bm h} \, {\bm m} \, ) \,
\biggl( \, {\bf I} +i \hbar \, \frac{1}{{\bf I} +{\bm h} \, {\bm m}}{\bm h} \, {\bf U} \, \biggr) \,,
\end{equation}
we obtain
\begin{equation}
\frac{1}{{\bf I} +{\bm h} \, {\bm m} +i \hbar \, {\bm h} \, {\bf U}}
= \frac{1}{{\bf I} +i \hbar \, {\bf H} \, {\bf U}} \,
\frac{1}{{\bf I} +{\bm h} \, {\bm m}} \,,
\end{equation}
where
\begin{equation}
{\bf H} = \frac{1}{{\bf I} +{\bm h} \, {\bm m}} \, {\bm h} \,.
\label{H-definition}
\end{equation}
The formula for correlation functions is then
\begin{equation}
\langle \, \Phi^{\otimes n} \, \rangle
= \pi_n \, \frac{1}{{\bf I} +i \hbar \, {\bf H} \, {\bf U}} \, {\bf P} \, {\bf 1}
\label{new-form-correlation}
\end{equation}
with
\begin{equation}
{\bf P} = \frac{1}{{\bf I} +{\bm h} \, {\bm m}} \, e^{{\bf \Phi}_\ast^{(0)}} \pi_0 \,.
\label{new-P}
\end{equation}
When the solution $\Phi_\ast^{(0)}$ satisfies the full equations of motion, we have
\begin{equation}
( \, {\bf Q} +{\bm m} \, ) \, e^{{\bf \Phi}_\ast^{(0)}} {\bf 1} = 0 \,.
\end{equation}
It then follows from~\eqref{Phi^(0)-coderivation} that
\begin{equation}
{\bm m} \, e^{{\bf \Phi}_\ast^{(0)}} {\bf 1} = 0 \,.
\end{equation}
In this case, ${\bf P}$ in~\eqref{new-P} therefore reduces to ${\bf P}^{(0)}$ in~\eqref{free-P}.
What is the interpretation of ${\bf P}$ in~\eqref{new-P}
when the solution $\Phi_\ast^{(0)}$ does not solve the full equations of motion?
First, we can show the relation
\begin{equation}
\triangle \, {\bf P} \, {\bf 1}
= {\bf P} \, {\bf 1} \otimes' {\bf P} \, {\bf 1}
\end{equation}
so that ${\bf P}$ can be written in the form
\begin{equation}
{\bf P} = e^{{\bf \Phi}_\ast} \, \pi_0
\label{Phi_ast}
\end{equation}
for an element $\Phi_\ast$ in $\mathcal{H}_1$,
where ${\bf \Phi}_\ast$ is the coderivation~\eqref{Phi-coderivation}
associated with $\Phi_\ast$.
Second, we can show that
\begin{equation}
{\bf M} \, {\bf P} = 0 \,.
\end{equation}
To prove this, let us first consider ${\bf Q} \, {\bf P}$.
Since ${\bf Q} \, {\bf P}^{(0)} = 0$, we have
\begin{equation}
{\bf Q} \, {\bf P}
= {\bf Q} \, \frac{1}{{\bf I} +{\bm h} \, {\bm m}} \, {\bf P}^{(0)}
= \biggl[ \, {\bf Q} \,,\, \frac{1}{{\bf I} +{\bm h} \, {\bm m}} \, \biggr] \, {\bf P}^{(0)} \,.
\end{equation}
The commutator in this expression can be calculated as follows:
\begin{equation}
\begin{split}
\biggl[ \, {\bf Q} \,,\, \frac{1}{{\bf I} +{\bm h} \, {\bm m}} \, \biggr]
& = {}-\frac{1}{{\bf I} +{\bm h} \, {\bm m}} \,
[ \, {\bf Q} \,,\, {\bf I} +{\bm h} \, {\bm m} \, ] \, 
\frac{1}{{\bf I} +{\bm h} \, {\bm m}} \\
& = {}-\frac{1}{{\bf I} +{\bm h} \, {\bm m}} \,
[ \, {\bf Q} \,,\, {\bm h} \, ] \, {\bm m} \,  
\frac{1}{{\bf I} +{\bm h} \, {\bm m}}
+\frac{1}{{\bf I} +{\bm h} \, {\bm m}} \,
{\bm h} \, [ \, {\bf Q} \,,\, {\bm m} \, ] \, 
\frac{1}{{\bf I} +{\bm h} \, {\bm m}} \\
& = {}-\frac{1}{{\bf I} +{\bm h} \, {\bm m}} \,
( \, {\bf I} -{\bf P}^{(0)} \, )  \, {\bm m} \,  
\frac{1}{{\bf I} +{\bm h} \, {\bm m}}
-\frac{1}{{\bf I} +{\bm h} \, {\bm m}} \,
{\bm h} \, {\bm m}^2 \, 
\frac{1}{{\bf I} +{\bm h} \, {\bm m}} \,.
\end{split}
\end{equation}
The coderivation ${\bm m}$ is annihilated by $\pi_0$ in ${\bf P}^{(0)}$:
\begin{equation}
{\bf P}^{(0)} \, {\bm m} = 0 \,.
\end{equation}
We thus have
\begin{equation}
\begin{split}
\biggl[ \, {\bf Q} \,,\, \frac{1}{{\bf I} +{\bm h} \, {\bm m}} \, \biggr]
& = {}-\frac{1}{{\bf I} +{\bm h} \, {\bm m}} \, {\bm m} \,  
\frac{1}{{\bf I} +{\bm h} \, {\bm m}}
-\frac{1}{{\bf I} +{\bm h} \, {\bm m}} \,
{\bm h} \, {\bm m}^2 \, 
\frac{1}{{\bf I} +{\bm h} \, {\bm m}} \\
& = {}-\frac{1}{{\bf I} +{\bm h} \, {\bm m}} \, 
( \, {\bf I} +{\bm h} \, {\bm m} \, ) \, {\bm m} \,  
\frac{1}{{\bf I} +{\bm h} \, {\bm m}}
= {}-{\bm m} \,  
\frac{1}{{\bf I} +{\bm h} \, {\bm m}} \,.
\end{split}
\end{equation}
We use this to write ${\bf Q} \, {\bf P}$ as
\begin{equation}
{\bf Q} \, {\bf P}
= {}-{\bm m} \,  
\frac{1}{{\bf I} +{\bm h} \, {\bm m}} \, {\bf P}^{(0)}
= {}-{\bm m} \, {\bf P} \,,
\end{equation}
and we conclude that ${\bf M} \, {\bf P} = 0$.
This implies that
$\Phi_\ast$ defined in~\eqref{Phi_ast} solves the full equations of motion.

Let us next consider the operator~${\bf H}$ in~\eqref{H-definition}.
We can show that ${\bf H}$ satisfies the following relations:
\begin{equation}
{\bf M} \, {\bf H} +{\bf H} \, {\bf M} = {\bf I} -{\bf P} \,, \qquad 
{\bf H} \, {\bf P} = 0 \,, \qquad
{\bf P} \, {\bf H} = 0 \,, \qquad
{\bf H}^2 = 0 \,.
\label{new-HK}
\end{equation}
The relations 
${\bf H} \, {\bf P} = 0$,
${\bf P} \, {\bf H} = 0$,
and ${\bf H}^2 = 0$ immediately follow from
${\bm h} \, {\bf P}^{(0)} = 0$,
${\bf P}^{(0)} \, {\bm h} = 0$,
and ${\bm h}^2 = 0$.
For the relation ${\bf M} \, {\bf H} +{\bf H} \, {\bf M} = {\bf I} -{\bf P}$,
we first consider ${\bf Q} \, {\bf H} +{\bf H} \, {\bf Q}$ 
and we write it as follows:
\begin{equation}
\begin{split}
{\bf Q} \, {\bf H} +{\bf H} \, {\bf Q}
& = \biggl[ \, {\bf Q} \,,\, \frac{1}{{\bf I} +{\bm h} \, {\bm m}} \, \biggr] \, {\bm h}
+\frac{1}{{\bf I} +{\bm h} \, {\bm m}} \, [ \, {\bf Q} \,,\, {\bm h} \, ] \\
& = {}-{\bm m} \, \frac{1}{{\bf I} +{\bm h} \, {\bm m}} \, {\bm h}
+\frac{1}{{\bf I} +{\bm h} \, {\bm m}} \, ( \, {\bf I} -{\bf P}^{(0)} \, ) \,.
\end{split}
\end{equation}
We therefore have
\begin{equation}
\begin{split}
{\bf M} \, {\bf H} +{\bf H} \, {\bf M}
& = \frac{1}{{\bf I} +{\bm h} \, {\bm m}} \, {\bm h} \, {\bm m}
+\frac{1}{{\bf I} +{\bm h} \, {\bm m}} \, ( \, {\bf I} -{\bf P}^{(0)} \, ) \\
& = \frac{1}{{\bf I} +{\bm h} \, {\bm m}} \, ( \, {\bf I} +{\bm h} \, {\bm m} \, )
-\frac{1}{{\bf I} +{\bm h} \, {\bm m}} \, {\bf P}^{(0)}
= {\bf I} -{\bf P} \,.
\end{split}
\end{equation}

\subsection{New form of the formula and its interpretation}
\label{interpretation-section}

Let us summarize the new form of the formula we obtained.
We denote the coderivation that describes the action by ${\bf M}$.
We then choose a solution $\Phi_\ast$ to the equations of motion
and we denote the associated coderivation by ${\bf \Phi}_\ast$:
\begin{equation}
{\bf M} \, {\bf P} = 0
\end{equation}
with
\begin{equation}
{\bf P} = e^{{\bf \Phi}_\ast} \, \pi_0 \,.
\end{equation}
The formula for correlation functions is given by
\begin{equation}
\langle \, \Phi^{\otimes n} \, \rangle
= \pi_n \, \frac{1}{{\bf I} +i \hbar \, {\bf H} \, {\bf U}} \, {\bf P} \, {\bf 1} \,,
\end{equation}
where ${\bf H}$ satisfies
\begin{equation}
{\bf M} \, {\bf H} +{\bf H} \, {\bf M} = {\bf I} -{\bf P} \,, \qquad 
{\bf H} \, {\bf P} = 0 \,, \qquad
{\bf P} \, {\bf H} = 0 \,, \qquad
{\bf H}^2 = 0 \,.
\label{H-conditions}
\end{equation}
This formula does not involve the division of the action into the free part and the interaction part.

The operator ${\bf H}$ satisfying the conditions~\eqref{H-conditions}
can be constructed in the same way as we did in subsection~\ref{nontrivial-vacuum-section}.
We define $\widetilde{\bf M}$ by
\begin{equation}
\widetilde{\bf M} = e^{-{\bf \Phi}_\ast} \, {\bf M} \, e^{{\bf \Phi}_\ast} \,.
\end{equation}
We then decompose $\pi_1 \, \widetilde{\bf M}$ as
\begin{equation}
\pi_ 1 \, \widetilde{\bf M}
= \widetilde{Q} \, \pi_1
+\sum_{n=2}^\infty \widetilde{m}_n \, \pi_n \,,
\end{equation}
and we define the coderivation $\widetilde{\bf Q}$
associated with $\widetilde{Q}$
and the coderivation $\widetilde{\bm m}_n$
associated with $\widetilde{m}_n$ for each $n$.
Note that $\pi_ 1 \, \widetilde{\bf M} \, \pi_0$ vanishes,
\begin{equation}
\pi_ 1 \, \widetilde{\bf M} \, \pi_0 = 0 \,,
\end{equation}
as $\Phi_\ast$ is a solution to the equations of motion.
We then have
\begin{equation}
\widetilde{\bf M}
= \widetilde{\bf Q} +\widetilde{\bm m}
\end{equation}
with
\begin{equation}
\widetilde{\bm m} = \sum_{n=2}^\infty \widetilde{\bm m}_n \,.
\end{equation}
We construct $\widetilde{h}$ satisfying
\begin{equation}
\widetilde{Q} \, \widetilde{h} +\widetilde{h} \, \widetilde{Q} = \mathbb{I} \,, \qquad
\widetilde{h}^2 = 0 \,,
\end{equation}
and $\widetilde{\bm h}$ satisfying
\begin{equation}
\widetilde{\bf Q} \, \widetilde{\bm h} +\widetilde{\bm h} \, \widetilde{\bf Q}
= {\bf I} -\pi_0 \,, \qquad
\widetilde{\bm h} \, \pi_0 = 0 \,, \qquad
\pi_0 \, \widetilde{\bm h} = 0 \,, \qquad
\widetilde{\bm h}^2 = 0 \,.
\end{equation}
The operator ${\bf H}$ is then given by
\begin{equation}
{\bf H} = e^{{\bf \Phi}_\ast} \,
\frac{1}{{\bf I} +\widetilde{\bm h} \, \widetilde{\bm m}} \, \widetilde{\bm h} \,
e^{-{\bf \Phi}_\ast} \,.
\end{equation}
The formula for correlation functions is
\begin{equation}
\langle \, \Phi^{\otimes n} \, \rangle
= \pi_n \, \frac{1}{{\bf I} +i \hbar \, {\bf H} \, {\bf U}} \, {\bf P} \, {\bf 1}
= \pi_n \, e^{{\bf \Phi}_\ast} \,
\frac{1}{{\bf I} +\widetilde{\bm h} \, \widetilde{\bm m}
+i \hbar \, \widetilde{\bm h} \, {\bf U}} \, {\bf 1} \,.
\end{equation}
While the construction of ${\bf H}$ is perturbative,
this does not imply that the resulting correlation functions are perturbative,
as we demonstrated in section~\ref{result-intro}.

We now have the formula for a given solution to the equations of motion.
What does the formula describe?
First, it reproduces perturbation theory around the solution $\Phi_\ast$.
Second, the solution $\Phi_\ast$ does not have to be real.
Third, when the inverse of ${\bf I} +i \hbar \, {\bf H} \, {\bf U}$ exists
nonperturbatively, the Schwinger-Dyson equations are satisfied.
Judging from these, we claim that the formula describes correlation functions
on the {\it Lefschetz thimble} associated with the solution~\cite{Witten:2010cx}.

To explain Lefschetz thimbles in the case of scalar field theories in zero dimensions,
let us replace the real variable $\varphi$ of the action $S$
by a complex variable $z$.
Consider a flow $z (t)$ parametrized by $t$ in the complex $z$ plane
which satisfies the downward flow equation:
\begin{equation}
\frac{dz}{dt} = i \, \frac{\partial \overline{S}}{\partial \bar{z}} \,, \qquad
\frac{d \bar{z}}{dt} = {}-i \, \frac{\partial S}{\partial z} \,.
\end{equation}
Along the flow, the imaginary part of $S$ increases as $t$ increases:
\begin{equation}
\frac{d \, {\rm Im} \, S}{dt}
= \frac{1}{2i} \, \biggl( \, \frac{d S}{dt} -\frac{d \overline{S}}{dt} \, \biggr)
= \frac{1}{2i} \, \biggl( \, \frac{\partial S}{\partial z} \, \frac{dz}{dt}
-\frac{\partial \overline{S}}{\partial \bar{z}} \, \frac{d \bar{z}}{dt} \, \biggr)
= \biggl| \, \frac{\partial S}{\partial z} \, \biggr|^2 > 0 \,.
\end{equation}
A Lefschetz thimble associated with a solution $z_\ast$ is defined
by a submanifold of the $z$ plane consisting of points
that can be reached at any $t$ by a flow that starts from $z_\ast$ at $t = -\infty$.
The path integral on a Lefschetz thimble is thus well defined.

Let us denote the Lefschetz thimble associated with $z_i$ by $\mathcal{J}_i$,
where $i$ labels solutions.
In general, the path integral over the real variable $\varphi$
should be understood as being defined by the path integral on $\mathcal{C}$ given by
\begin{equation}
\mathcal{C} = \sum_i n_i \, \mathcal{J}_i \,,
\end{equation}
where $n_i$'s are integers and there is a procedure to determine $n_i$~\cite{Witten:2010cx}.

\begin{figure}
\begin{center}
\includegraphics[width=6cm]{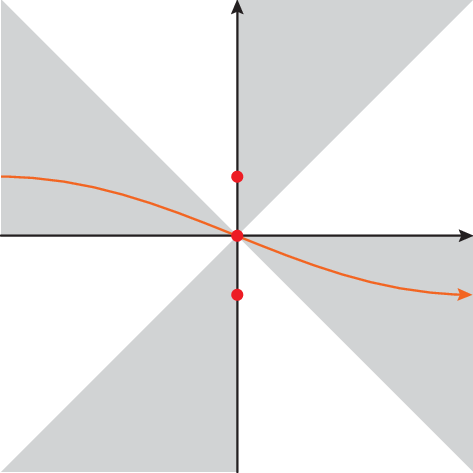}
\end{center}
\caption{The integration contour on the Lefschetz thimble associated with the solution
$\varphi=0$ to the equation of motion of $\varphi^4$ theory.
The integral converges when the contour extends to the shaded regions
as $| \varphi | \to \ \infty \,$.
}
\label{figure-phi^4-thimble}
\end{figure}

In section~\ref{result-intro}, we considered the action
\begin{equation}
S = {}-\frac{1}{2} \, m^2 \, \varphi^2 -\frac{1}{4} \, \lambda \, \varphi^4
\end{equation}
in the Lorentzian case.
The solutions to the equation of motion are
\begin{equation}
\varphi = 0 \,, \pm \frac{i m}{\sqrt{\lambda}} \,,
\end{equation}
which are depicted as red dots in figure~\ref{figure-phi^4-thimble}.
In this case,
only the Lefschetz thimble associated with the trivial solution $\varphi = 0$ contributes,
and that is why correlation functions were reproduced in subsection~\ref{Lorentzian-quartic}.
We will present evidence that supports our claim
for more nontrivial cases in the next section.

\section{Evidence for the claim}
\setcounter{equation}{0}

\subsection{Airy function}

Consider the action given by
\begin{equation}
S = {}-a \, \varphi -\frac{1}{3} \, \varphi^3 \,,
\end{equation}
where $a$ is a real constant.
The partition function is given by
\begin{equation}
Z = \int_{-\infty}^\infty d \varphi \, e^{\frac{i}{\hbar} S} \,.
\end{equation}
When we set $\hbar = 1$,
this is expressed in terms of the Airy function of the first kind ${\rm Ai} \, (a)$ as follows:\footnote{
Our definition of ${\rm Ai} \, (a)$ is the same as that of ${\rm AiryAi} \, [a]$ in {\it Mathematica}.
}
\begin{equation}
Z =  2 \pi \, {\rm Ai} \, (a) \,.
\end{equation}
The correlation functions are defined by
\begin{equation}
\langle \, \varphi^n \, \rangle
= \frac{1}{Z} \int_{-\infty}^\infty d \varphi \, \varphi^n \, e^{\frac{i}{\hbar} S} \,.
\end{equation}
They can be calculated from the partition function as follows:
\begin{equation}
\langle \, \varphi^n \, \rangle
= ( \, i \hbar \, )^n \frac{1}{Z} \, \frac{d^n Z}{da^n} \,.
\label{Airy-derivatives}
\end{equation}
The equation of motion is given by
\begin{equation}
\varphi^2 +a = 0 \,.
\end{equation}

\subsubsection{The case where $a > 0$}

\begin{figure}[t]
\centering\includegraphics[width=6cm]{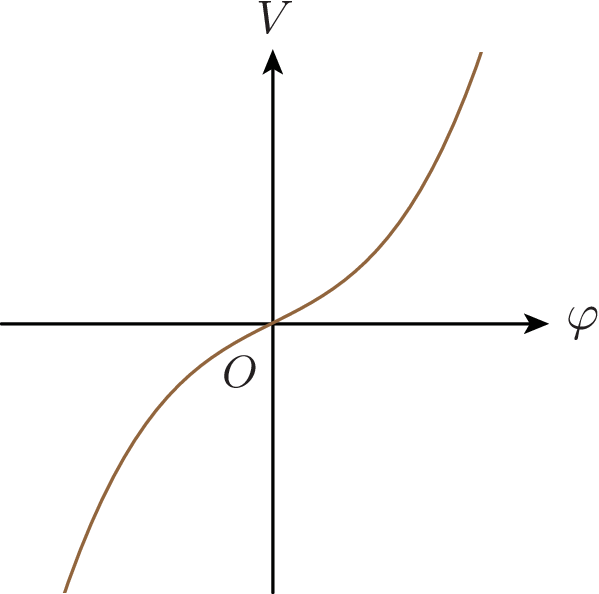}
\caption{The potential in the case where $a > 0$.}
\label{figure-potential-Airy-positive}
\end{figure}

\begin{figure}[t]
\centering\includegraphics[width=7cm]{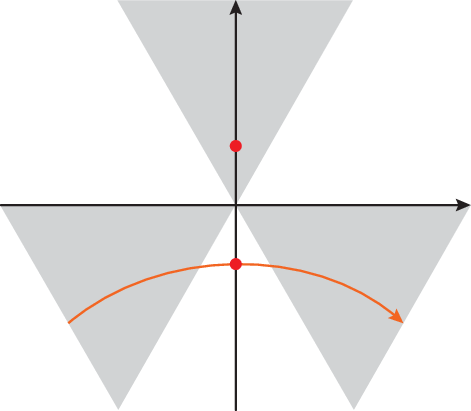}
\caption{The integration contour on the Lefschetz thimble associated with the solution $\varphi = {}-i\sqrt{a}$.}
\label{figure-airy1}
\end{figure}

The potential when $a > 0$ is shown in figure~\ref{figure-potential-Airy-positive}.
There are no real solutions to the equation of motion,
and the complex solutions are
\begin{equation}
\varphi = \pm i \sqrt{a} \,.
\end{equation}
It is known that
only the Lefschetz thimble associated with the solution $\varphi = {}-i\sqrt{a}$ contributes
in this case.
Namely, this is an example where the theory consists of a single Lefschetz thimble
associated with a nontrivial solution.
The integration contour on the Lefschetz thimble
associated with the solution $\varphi = {}-i\sqrt{a}$ is shown in figure~\ref{figure-airy1}.

Let us calculate correlation functions following the general procedure.
We expand $\varphi$ as
\begin{equation}
\varphi = {}-i \sqrt{a} +\widetilde{\varphi} \,.
\end{equation}
The action in terms of $\widetilde{\varphi}$ is
\begin{equation}
S = \frac{2i}{3} \, a \sqrt{a} +i \sqrt{a} \, \widetilde{\varphi}^{\, 2}
-\frac{1}{3} \, \widetilde{\varphi}^{\, 3} \,.
\end{equation}
The solution $\Phi_\ast$ is
\begin{equation}
\Phi_\ast = {}-i \sqrt{a} \, c \,,
\end{equation}
and the associated coderivation ${\bf \Phi}_\ast$ is
\begin{equation}
{\bf \Phi}_\ast \, {\bf 1} = {}-i \sqrt{a} \, c \,.
\end{equation}
The operators $\widetilde{Q}$ and $\widetilde{m}_2$ are given by
\begin{equation}
\widetilde{Q} \, c = {}-2i \sqrt{a} \, d \,, \qquad
\widetilde{m}_{2} \, ( \, c \otimes c \, ) = d \,.
\end{equation}
The contracting homotopy $\widetilde{h}$ is
\begin{equation}
\widetilde{h} \, d = \frac{i}{2\sqrt{a}} \, c \,.
\end{equation}
The formula for correlation functions is
\begin{equation}
\langle \, \Phi^{\otimes n} \, \rangle
= \pi_n \, e^{{\bf \Phi}_\ast} \widetilde{\bm f} \, {\bf 1} \,,
\end{equation}
where
\begin{equation}
\widetilde{\bm f}
= \frac{1}{{\bf I} +\widetilde{\bm h} \, \widetilde{\bm m}_2
+ i\hbar \, \widetilde{\bm h} \, {\bf U}} \,.
\label{Airy-positive-f-tilde}
\end{equation}
The matrix forms of the operators $\widetilde{\bm h} \, \widetilde{\bm m}_2$
and $\widetilde{\bm h} \, {\bf U}$ are
\begin{equation}
\widetilde{\bm h} \, \widetilde{\bm m}_2
= \frac{i}{2\sqrt{a}} \, \left(
\begin{array}{ccccccc}
0 & 0 & 0 & 0 & 0 & 0 & \ldots \\
0 & 0 & 1 & 0 & 0 & 0 & \ldots \\
0 & 0 & 0 & 1 & 0 & 0 & \ldots \\
0 & 0 & 0 & 0 & 1 & 0 & \ldots \\
0 & 0 & 0 & 0 & 0 & 1 & \ldots \\
0 & 0 & 0 & 0 & 0 & 0 & \ldots \\
\vdots & \vdots & \vdots & \vdots & \vdots & \vdots & \ddots
\end{array}
\right) \,, \quad
\widetilde{\bm h} \, {\bf U}
= \frac{i}{2\sqrt{a}} \, \left(
\begin{array}{ccccccc}
0 & 0 & 0 & 0 & 0 & 0 & \ldots \\
0 & 0 & 0 & 0 & 0 & 0 & \ldots \\
1 & 0 & 0 & 0 & 0 & 0 & \ldots \\
0 & 2 & 0 & 0 & 0 & 0 & \ldots \\
0 & 0 & 3 & 0 & 0 & 0 & \ldots \\
0 & 0 & 0 & 4 & 0 & 0 & \ldots \\
\vdots & \vdots & \vdots & \vdots & \vdots & \vdots & \ddots
\end{array}
\right) \,.
\end{equation}
The $n$-point functions with $n \le 3$ are given by
\begin{align}
\langle \, \varphi \, \rangle & = {}-i \sqrt{a} +\widetilde{\bm f}_{10} \,, \\
\langle \, \varphi^2 \, \rangle
& = {}-a -2i \sqrt{a} \, \widetilde{\bm f}_{10} +\widetilde{\bm f}_{20} \,,
\label{Airy-positive-two-point}
\\
\langle \, \varphi^3 \, \rangle
& = i a \sqrt{a}
-3 a \, \widetilde{\bm f}_{10}
-3i \sqrt{a} \, \widetilde{\bm f}_{20} +\widetilde{\bm f}_{30} \,.
\end{align}
We calculate the right-hand side of each equation
using~\eqref{Airy-positive-f-tilde}
and compare it with the left-hand side calculated from~\eqref{Airy-derivatives}.
We set $\hbar = 1$ when we confirm these relations.

Let us begin with the one-point function.
For $a=1$, we calculate $\widetilde{\bm f}_{10}$ with $N=50$. We find
\begin{equation}
\begin{split}
\langle \, \varphi \, \rangle
& \simeq -1.17632196714 \, i \,, \\
{}-i \sqrt{a} +\widetilde{\bm f}_{10}
& \simeq -1.17632196731 \, i \,.
\end{split}
\end{equation}
For $a=2$, we again calculate $\widetilde{\bm f}_{10}$ with $N=50$ and find
\begin{equation}
\begin{split}
\langle \, \varphi \, \rangle
& \simeq -1.5201633881848286 \, i \,, \\
{}-i \sqrt{a} +\widetilde{\bm f}_{10}
& \simeq -1.5201633881848252 \, i \,.
\end{split}
\end{equation}
The coupling is weaker when $a$ is larger,
and we have found that the agreement is better for $a = 2$.
Let us consider the case where the coupling is stronger.
For $a=0.1$, we calculate $\widetilde{\bm f}_{10}$ with $N=50$ and find
\begin{equation}
\begin{split}
\langle \, \varphi \, \rangle
& \simeq -0.7811 \, i \,, \\
{}-i \sqrt{a} +\widetilde{\bm f}_{10}
& \simeq -0.7822 \, i \,.
\end{split}
\end{equation}
The agreement is not so good, but it improves in the calculation with $N=100$:
\begin{equation}
\begin{split}
\langle \, \varphi \, \rangle
& \simeq -0.781069 \, i \,, \\
{}-i \sqrt{a} +\widetilde{\bm f}_{10}
& \simeq -0.781005 \, i \,.
\end{split}
\end{equation}

The two-point function can be analytically obtained from the Schwinger-Dyson equations. We find
\begin{equation}
\langle \, \varphi^2 \, \rangle = {}-a \,.
\end{equation}
We calculate $\widetilde{\bm f}_{10}$ and $\widetilde{\bm f}_{20}$ with $N=50$
and find
\begin{equation}
-2i \sqrt{a} \, \widetilde{\bm f}_{10} +\widetilde{\bm f}_{20} = 0
\end{equation}
for any $a$, which is consistent with~\eqref{Airy-positive-two-point}.

The three-point function is related to the one-point function
via the Schwinger-Dyson equations.
For $a=1$, we calculate $\widetilde{\bm f}_{10}$, $\widetilde{\bm f}_{20}$,
and $\widetilde{\bm f}_{30}$ with $N=50$ and find
\begin{equation}
\begin{split}
\langle \, \varphi^3 \, \rangle
& \simeq 0.17632196714 \, i \,, \\
i a \sqrt{a}
-3 a \, \widetilde{\bm f}_{10}
-3i \sqrt{a} \, \widetilde{\bm f}_{20} +\widetilde{\bm f}_{30}
& \simeq 0.17632196731 \, i \,.
\end{split}
\end{equation}

We conclude that the formula based on quantum $A_\infty$ algebras
reproduces the correlation functions in the path integral formalism
with high precision in all cases.
We consider this as nontrivial evidence
for our claim when the theory consists of a single Lefschetz thimble
associated with a nontrivial solution.

\subsubsection{The case where $a < 0$}

\begin{figure}
\centering\includegraphics[width=6cm]{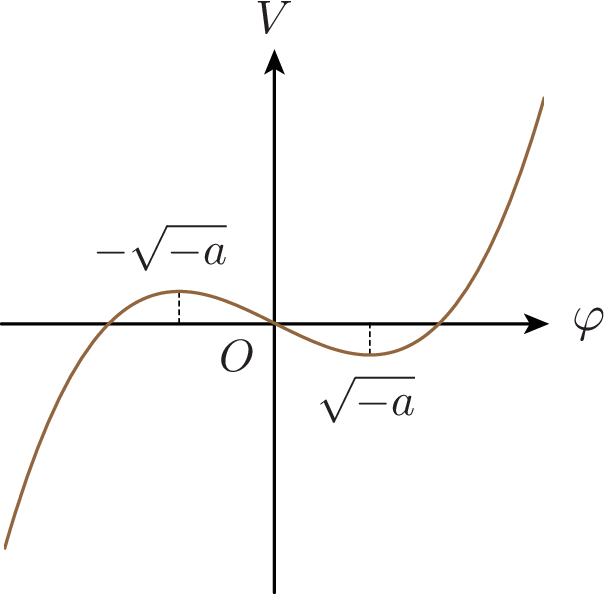}
\caption{The potential in the case where $a < 0$.}
\label{figure-potential-Airy-negative}
\end{figure}

\begin{figure}
\begin{center}
\includegraphics[width=7cm]{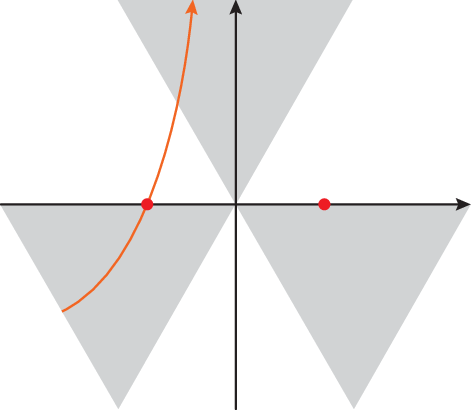}
\hspace{10mm}
\includegraphics[width=7cm]{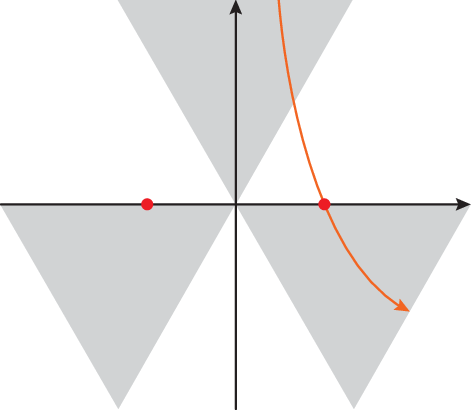}
\end{center}
\caption{
The left figure shows the integration contour
on the Lefschetz thimble associated with the solution $\varphi = {}-\sqrt{-a}$.
The right figure shows the integration contour
on the Lefschetz thimble associated with the solution $\varphi = \sqrt{-a}$.}
\label{figure-airy2}
\end{figure}

The potential when $a < 0$ is shown in figure~\ref{figure-potential-Airy-negative}.
The solutions to the equation of motion are
\begin{equation}
\varphi = \pm \sqrt{-a} \,.
\end{equation}
It is known that
both of the two Lefeschetz thimbles associated with the solutions
$\varphi = {}-\sqrt{-a}$ and $\varphi = \sqrt{-a}$ contribute
in this case.
See figure~\ref{figure-airy2} for the integration contours on these Lefschetz thimbles.

We denote the partition function associated with the solution $\varphi = {}-\sqrt{-a}$
by $Z_{-}$
and the partition function associated with the solution $\varphi = \sqrt{-a}$
by $Z_{+}$.
By evaluating the integrals on the contours in figure~\ref{figure-airy2},
we find after setting $\hbar =1$ that
\begin{equation}
Z_{-} = \pi \, ( \, {\rm Ai} \, (a) +i \, {\rm Bi} \, (a) \, ) \,, \qquad
Z_{+} = \pi \, ( \, {\rm Ai} \, (a) -i \, {\rm Bi} \, (a) \, ) \,,
\end{equation}
where ${\rm Bi} \, (a)$ is the Airy function of the second kind.\footnote{
Our definition of ${\rm Bi} \, (a)$ is the same as that of ${\rm AiryBi} \, [a]$ in {\it Mathematica}.
}
We denote the $n$-point function of the theory $Z_{-}$ by $\langle \, \varphi^n \, \rangle_{-}$
and the $n$-point function of the theory $Z_{+}$ by $\langle \, \varphi^n \, \rangle_{+}$.
The partition function $Z$ of the full theory is
\begin{equation}
Z = Z_{-} +Z_{+} = 2 \pi \, {\rm Ai} \, (a) \,,
\end{equation}
and the $n$-point function of the full theory $\langle \, \varphi^n \, \rangle$ is given by
\begin{equation}
\langle \, \varphi^n \, \rangle
= \frac{Z_{-}}{Z_{-} +Z_{+}} \, \langle \, \varphi^n \, \rangle_{-}
+\frac{Z_{+}}{Z_{-} +Z_{+}} \, \langle \, \varphi^n \, \rangle_{+} \,.
\end{equation}

Let us calculate the correlation functions from the formula
based on quantum $A_\infty$ algebras.
We expand $\varphi$ as
\begin{equation}
\varphi = \pm \sqrt{-a} +\widetilde{\varphi} \,.
\end{equation}
The action in terms of $\widetilde{\varphi}$ is
\begin{equation}
S = {}\mp \frac{2}{3} \, a \sqrt{-a}
\mp \sqrt{-a} \, \widetilde{\varphi}^{\, 2}
-\frac{1}{3} \, \widetilde{\varphi}^{\, 3} \,.
\end{equation}
The solution $\Phi_\ast$ is
\begin{equation}
\Phi_\ast = {}\pm \sqrt{-a} \, c \,,
\end{equation}
and the associated coderivation ${\bf \Phi}_\ast$ is
\begin{equation}
{\bf \Phi}_\ast \, {\bf 1} = {}\pm \sqrt{-a} \, c \,.
\end{equation}
The operators $\widetilde{Q}$ and $\widetilde{m}_2$ are given by
\begin{equation}
\widetilde{Q} \, c = {}\pm 2 \sqrt{-a} \, d \,, \qquad
\widetilde{m}_{2} \, ( \, c \otimes c \, ) = d \,,
\end{equation}
and the contracting homotopy $\widetilde{h}$ is
\begin{equation}
\widetilde{h} \, d = {}\pm \frac{1}{2 \sqrt{-a}} \, c \,.
\end{equation}
The formula for correlation functions is
\begin{equation}
\langle \, \Phi^{\otimes n} \, \rangle_{\pm}
= \pi_n \, e^{{\bf \Phi}_\ast} \widetilde{\bm f} \, {\bf 1} \,,
\end{equation}
where
\begin{equation}
\widetilde{\bm f}
= \frac{1}{{\bf I} +\widetilde{\bm h} \, \widetilde{\bm m}_2
+ i\hbar \, \widetilde{\bm h} \, {\bf U}} \,.
\end{equation}
The matrix forms of the operators $\widetilde{\bm h} \, \widetilde{\bm m}_2$
and $\widetilde{\bm h} \, {\bf U}$ are
\begin{equation}
\widetilde{\bm h} \, \widetilde{\bm m}_2
= {}\pm \frac{1}{2 \sqrt{-a}} \, \left(
\begin{array}{ccccccc}
0 & 0 & 0 & 0 & 0 & 0 & \ldots \\
0 & 0 & 1 & 0 & 0 & 0 & \ldots \\
0 & 0 & 0 & 1 & 0 & 0 & \ldots \\
0 & 0 & 0 & 0 & 1 & 0 & \ldots \\
0 & 0 & 0 & 0 & 0 & 1 & \ldots \\
0 & 0 & 0 & 0 & 0 & 0 & \ldots \\
\vdots & \vdots & \vdots & \vdots & \vdots & \vdots & \ddots
\end{array}
\right) \,, \quad
\widetilde{\bm h} \, {\bf U}
= {}\pm \frac{1}{2 \sqrt{-a}} \, \left(
\begin{array}{ccccccc}
0 & 0 & 0 & 0 & 0 & 0 & \ldots \\
0 & 0 & 0 & 0 & 0 & 0 & \ldots \\
1 & 0 & 0 & 0 & 0 & 0 & \ldots \\
0 & 2 & 0 & 0 & 0 & 0 & \ldots \\
0 & 0 & 3 & 0 & 0 & 0 & \ldots \\
0 & 0 & 0 & 4 & 0 & 0 & \ldots \\
\vdots & \vdots & \vdots & \vdots & \vdots & \vdots & \ddots
\end{array}
\right) \,.
\end{equation}
The $n$-point functions with $n \le 3$ are given by
\begin{align}
\langle \, \varphi \, \rangle_{\pm} & = {}\pm \sqrt{-a} +\widetilde{\bm f}_{10} \,, \\
\langle \, \varphi^2 \, \rangle_{\pm}
& = {}-a \pm 2 \sqrt{-a} \, \widetilde{\bm f}_{10} +\widetilde{\bm f}_{20} \,, \\
\langle \, \varphi^3 \, \rangle_{\pm}
& = {}\mp a \sqrt{-a}
-3 a \, \widetilde{\bm f}_{10}
\pm 3 \sqrt{-a} \, \widetilde{\bm f}_{20} +\widetilde{\bm f}_{30} \,.
\end{align}

Let us calculate the one-point function and the three-point function
for each Lefschetz thimble.
We set $\hbar = 1$.
For the Lefschetz thimble associated with the solution $\varphi = {}-\sqrt{-a}$,
we find for $a = -1$ with $N=50$ that
\begin{equation}
\begin{split}
\langle \, \varphi \, \rangle_{-}
& \simeq -1.06944263 +0.18869689 \, i \,, \\
{}-\sqrt{-a} +\widetilde{\bm f}_{10}
& \simeq -1.06944243 +0.18869651 \, i
\end{split}
\end{equation}
and
\begin{equation}
\begin{split}
\langle \, \varphi^3 \, \rangle_{-}
& \simeq -1.06944263 -0.81130311 \, i \,, \\
a \sqrt{-a}
-3 a \, \widetilde{\bm f}_{10}
-3 \sqrt{-a} \, \widetilde{\bm f}_{20} +\widetilde{\bm f}_{30}
& \simeq -1.06944243 -0.81130349 \, i \,.
\end{split}
\end{equation}
For the Lefschetz thimble associated with the solution $\varphi = \sqrt{-a}$,
we find for $a = -1$ with $N=50$ that
\begin{equation}
\begin{split}
\langle \, \varphi \, \rangle_{+}
& \simeq 1.06944263 + 0.18869689 \, i \,, \\
\sqrt{-a} +\widetilde{\bm f}_{10}
& \simeq 1.06944243 + 0.18869651 \, i
\end{split}
\end{equation}
and
\begin{equation}
\begin{split}
\langle \, \varphi^3 \, \rangle_{+}
& \simeq 1.06944263 -0.81130311 \, i \,, \\
{}-a \sqrt{-a}
-3 a \, \widetilde{\bm f}_{10}
+3 \sqrt{-a} \, \widetilde{\bm f}_{20} +\widetilde{\bm f}_{30}
& \simeq 1.06944243 -0.81130349 \, i \,.
\end{split}
\end{equation}

The correlation functions in the path integral formalism
are again reproduced from the formula based on quantum $A_\infty$ algebras
with high precision in all cases.
This is an example where there are multiple real solutions
to the equation of motion,
and we found that the formula gives different results
for different solutions.
The results are consistent with our claim
that the formula gives the correlation functions
on the Lefschetz thimble associated with the solution we chose.

Since the correlation functions on each Lefschetz thimble are reproduced,
the correlation functions of the full theory given by
\begin{equation}
\langle \, \varphi^n \, \rangle
= \frac{Z_{-}}{Z_{-} +Z_{+}} \, \langle \, \varphi^n \, \rangle_{-}
+\frac{Z_{+}}{Z_{-} +Z_{+}} \, \langle \, \varphi^n \, \rangle_{+}
\end{equation}
are reproduced if we know the ratio of the partition functions $Z_{-}$ and $Z_{+}$.
We will describe the method of expressing ratios of partition functions
in terms of quantum $A_\infty$ algebras in a forthcoming paper~\cite{Konosu:2024}.

\subsection{The double-well potential}

\begin{figure}[t]
\centering\includegraphics[width=6cm]{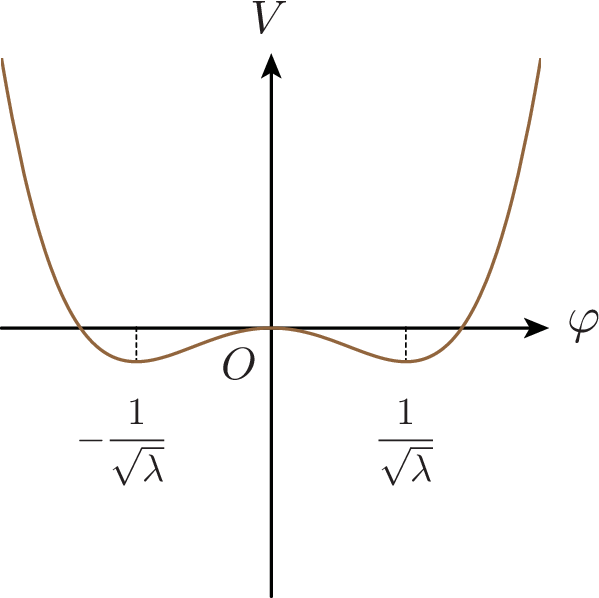}
\caption{The double-well potential.}
\label{figure-double-well-potential}
\end{figure}

\begin{figure}[t]
\centering\includegraphics[width=6cm]{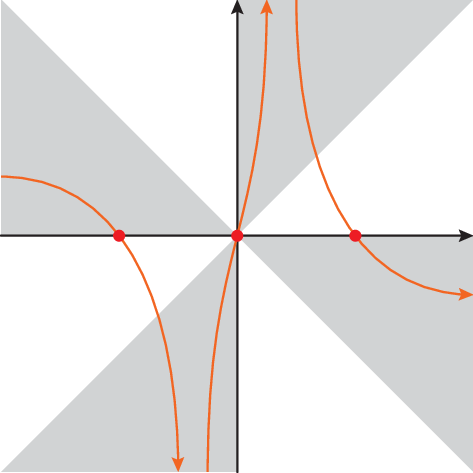}
\caption{The integration contours
on the Lefschetz thimbles associated with the three solutions.}
\label{figure-double-well2}
\end{figure}

In subsection~\ref{Lorentzian-quartic},
we considered $\varphi^4$ theory with the action given by
\begin{equation}
S = {}-\frac{1}{2} \, m^{2} \, \varphi^2 -\frac{1}{4} \lambda \, \varphi^4 \,.
\end{equation}
Let us consider the theory with the double-well potential.
The action is given by
\begin{equation}
S = \frac{1}{2} \, m^{2} \, \varphi^2 -\frac{1}{4} \lambda \, \varphi^4 \,.
\end{equation}
See figure~\ref{figure-double-well-potential} for the shape of the potential.
We set $m^2=1$.
The equation of motion is given by
\begin{equation}
\varphi -\lambda \, \varphi^3 = 0 \,,
\end{equation}
and the solutions are
\begin{equation}
\varphi = 0 \,,\, \pm\frac{1}{\sqrt{\lambda}} \,.
\end{equation}
All of the three Lefschetz thimbles associated with the three solutions
contribute in this case.
See figure~\ref{figure-double-well2} for the integration contours on these Lefschetz thimbles.

We denote the partition function associated with the solution $\varphi = -1/\sqrt{\lambda}$
by $Z_{-}$,
the partition function associated with the solution $\varphi = 0$
by $Z_{0}$,
and the partition function associated with the solution $\varphi = 1/\sqrt{\lambda}$
by $Z_{+}$.
For the $n$-point function,
we denote the $n$-point function of the theory $Z_{-}$ by $\langle \, \varphi^n \, \rangle_{-}$,
the $n$-point function of the theory $Z_{0}$ by $\langle \, \varphi^n \, \rangle_{0}$,
and the $n$-point function of the theory $Z_{+}$ by $\langle \, \varphi^n \, \rangle_{+}$.
The partition function $Z$ of the full theory is
\begin{equation}
Z = Z_{-} +Z_{0} +Z_{+} \,,
\end{equation}
and the $n$-point function of the full theory $\langle \, \varphi^n \, \rangle$ is given by
\begin{equation}
\langle \, \varphi^n \, \rangle
= \frac{Z_{-}}{Z_{-} +Z_{0} +Z_{+}} \, \langle \, \varphi^n \, \rangle_{-}
+\frac{Z_{0}}{Z_{-} +Z_{0} +Z_{+}} \, \langle \, \varphi^n \, \rangle_{0}
+\frac{Z_{+}}{Z_{-} +Z_{0} +Z_{+}} \, \langle \, \varphi^n \, \rangle_{+} \,.
\end{equation}

Let us calculate the correlation functions from the formula
based on quantum $A_\infty$ algebras for each Lefschetz thimble.
For the Lefschetz thimble associated with the solution $\varphi = 0$,
we do not need to shift the original field $\varphi$.
The operator $Q$ and $m_{3}$ are given by
\begin{equation}
Q \, c = {}-d \,, \qquad
m_3 \, ( \, c \otimes c \otimes c \, ) = \lambda \, d \,,
\end{equation}
and the contracting homotopy $h$ is given by
\begin{equation}
h \, d = {}-c \,.
\end{equation}
Then the correlation functions are given by
\begin{equation}
\langle \, \Phi^{\otimes} \, \rangle_{0} = \pi_{n} \, {\bm f} \, {\bf 1} \,,
\end{equation}
where
\begin{equation}
{\bm f} = \frac{1}{{\bf I} +{\bm h} \, {\bm m}_{3} +i \hbar \, {\bm h} \, {\bf U}} \,.
\end{equation}
The matrix forms of the operators ${\bm h} \, {\bm m}_3$
and ${\bm h} \, {\bf U}$ are
\begin{equation}
{\bm h} \, {\bm m}_3 = {}-\lambda \, \left(
\begin{array}{ccccccc}
0 & 0 & 0 & 0 & 0 & 0 & \ldots \\
0 & 0 & 0 & 1 & 0 & 0 & \ldots \\
0 & 0 & 0 & 0 & 1 & 0 & \ldots \\
0 & 0 & 0 & 0 & 0 & 1 & \ldots \\
0 & 0 & 0 & 0 & 0 & 0 & \ldots \\
0 & 0 & 0 & 0 & 0 & 0 & \ldots \\
\vdots & \vdots & \vdots & \vdots & \vdots & \vdots & \ddots
\end{array}
\right) \,, \quad
{\bm h} \, {\bf U}
= {}- \, \left(
\begin{array}{ccccccc}
0 & 0 & 0 & 0 & 0 & 0 & \ldots \\
0 & 0 & 0 & 0 & 0 & 0 & \ldots \\
1 & 0 & 0 & 0 & 0 & 0 & \ldots \\
0 & 2 & 0 & 0 & 0 & 0 & \ldots \\
0 & 0 & 3 & 0 & 0 & 0 & \ldots \\
0 & 0 & 0 & 4 & 0 & 0 & \ldots \\
\vdots & \vdots & \vdots & \vdots & \vdots & \vdots & \ddots
\end{array}
\right) \,,
\end{equation}
and the $n$-point function is given by
\begin{equation}
\langle \, \varphi^{n} \, \rangle_0 = {\bm f}_{n0} \,.
\end{equation}
We set $\hbar = 1$ when we confirm this relation in what follows.

The one-point function $\langle\,\varphi\,\rangle_0$ vanishes
because of the $\mathbb{Z}_2$ symmetry:
\begin{equation}
\langle \, \varphi \, \rangle_{0} = 0 \,.
\end{equation}
We calculate ${\bm f}_{10}$ with $N = 100$ to find
\begin{equation}
{\bm f}_{10} = 0
\end{equation}
for any $\lambda$, which is consistent with $\langle \, \varphi \, \rangle_{0} = 0$.
For the two-point function $\langle \, \varphi^2 \, \rangle_{0}$,
we choose $\lambda = 1$ and calculate ${\bm f}_{20}$ with $N = 100$ to find
\begin{equation}
\begin{split}
\langle \, \varphi^{2} \, \rangle_{0}
& \simeq -0.2598 + 0.4376 \, i \,, \\
{\bm f}_{20}
& \simeq -0.2608 + 0.4391 \, i \,.
\end{split}
\end{equation}

Let us next consider the Lefschetz thimbles associated with the solutions $\varphi=\pm 1/\sqrt{\lambda}$.
We expand $\varphi$ as
\begin{equation}
\varphi = \pm \frac{1}{\sqrt{\lambda}} +\widetilde{\varphi} \,.
\end{equation}
The action in terms of $\widetilde{\varphi}$ is given by
\begin{equation}
S = \frac{1}{4 \lambda} -\widetilde{\varphi}^{\,2}
\mp \sqrt{\lambda} \, \widetilde{\varphi}^{\,3}
-\frac{1}{4} \, \lambda \, \widetilde{\varphi}^{\, 4} \,.
\end{equation}
The solution $\Phi_\ast$ is
\begin{equation}
\Phi_\ast = {}\pm \frac{1}{\sqrt{\lambda}}\, c \,,
\end{equation}
and the associated coderivation ${\bf \Phi}_\ast$ is
\begin{equation}
{\bf \Phi}_\ast \, {\bf 1} = {}\pm \frac{1}{\sqrt{\lambda}} \, c \,.
\end{equation}
The operators $\widetilde{Q}$, $\widetilde{m}_{2}$, and $\widetilde{m}_{3}$ are given by
\begin{equation}
\widetilde{Q} \, c  = 2 \, d \,, \qquad
\widetilde{m}_2 \, ( \, c \otimes c \, ) = {}\pm 3 \, \sqrt{\lambda} \, d \,, \qquad
\widetilde{m}_3 \, ( \, c \otimes c \otimes c \, ) = \lambda \, d \,,
\end{equation}
and the contracting homotopy $\widetilde{h}$ is given by
\begin{equation}
\widetilde{h} \, d = \frac{1}{2} \, c \,.
\end{equation}
The formula for correlation functions is
\begin{equation}
\langle \, \Phi^{\otimes n} \, \rangle_{\pm}
= \pi_n \, e^{{\bf \Phi}_\ast} \widetilde{\bm f} \, {\bf 1} \,,
\end{equation}
where
\begin{equation}
\widetilde{\bm f}
= \frac{1}{{\bf I} +\widetilde{\bm h} \, \widetilde{\bm m}_2+\widetilde{\bm h} \, \widetilde{\bm m}_3
+ i\hbar \, \widetilde{\bm h} \, {\bf U}} \,.
\end{equation}
The matrix forms of the operators $\widetilde{\bm h} \, \widetilde{\bm m}_2,\,\widetilde{\bm h} \, \widetilde{\bm m}_3$
and $\widetilde{\bm h} \, {\bf U}$ are
\begin{align}
\widetilde{\bm h} \, \widetilde{\bm m}_2
= {}\pm \frac{3\sqrt{\lambda}}{2}  \, &\left(
\begin{array}{ccccccc}
0 & 0 & 0 & 0 & 0 & 0 & \ldots \\
0 & 0 & 1 & 0 & 0 & 0 & \ldots \\
0 & 0 & 0 & 1 & 0 & 0 & \ldots \\
0 & 0 & 0 & 0 & 1 & 0 & \ldots \\
0 & 0 & 0 & 0 & 0 & 1 & \ldots \\
0 & 0 & 0 & 0 & 0 & 0 & \ldots \\
\vdots & \vdots & \vdots & \vdots & \vdots & \vdots & \ddots
\end{array}
\right) \,, \quad
\widetilde{\bm h} \, \widetilde{\bm m}_3
= {} \frac{\lambda}{2} \,
\left(
\begin{array}{ccccccc}
0 & 0 & 0 & 0 & 0 & 0 & \ldots \\
0 & 0 & 0 & 1 & 0 & 0 & \ldots \\
0 & 0 & 0 & 0 & 1 & 0 & \ldots \\
0 & 0 & 0 & 0 & 0 & 1 & \ldots \\
0 & 0 & 0 & 0 & 0 & 0 & \ldots \\
0 & 0 & 0 & 0 & 0 & 0 & \ldots \\
\vdots & \vdots & \vdots & \vdots & \vdots & \vdots & \ddots
\end{array}
\right)\,,
\nonumber \\
\widetilde{\bm h} \, {\bf U}
= {} \frac{1}{2} \, &\left(
\begin{array}{ccccccc}
0 & 0 & 0 & 0 & 0 & 0 & \ldots \\
0 & 0 & 0 & 0 & 0 & 0 & \ldots \\
1 & 0 & 0 & 0 & 0 & 0 & \ldots \\
0 & 2 & 0 & 0 & 0 & 0 & \ldots \\
0 & 0 & 3 & 0 & 0 & 0 & \ldots \\
0 & 0 & 0 & 4 & 0 & 0 & \ldots \\
\vdots & \vdots & \vdots & \vdots & \vdots & \vdots & \ddots
\end{array}
\right)\,.
\end{align}

The one-point function and the two-point function are given by
\begin{align}
\langle \, \varphi \, \rangle_{\pm} & = {}\pm \frac{1}{\sqrt{\lambda}} +\widetilde{\bm f}_{10} \,, \\
\langle \, \varphi^2 \, \rangle_{\pm}
& = {}\frac{1}{\lambda} \pm \frac{2}{\sqrt{\lambda}} \, \widetilde{\bm f}_{10} +\widetilde{\bm f}_{20} \,.
\end{align}
We set $\hbar = 1$ when we confirm these relations in what follows.
For the Lefschetz thimble associated with the solution $\varphi = 1/\sqrt{\lambda}$,
we find for $\lambda = 1$ with $N=30$ that
\begin{equation}
\begin{split}
\langle \, \varphi \, \rangle_{+}
& \simeq 1.204933 + 0.26012 \,i \,, \\
\frac{1}{\sqrt{\lambda}} +\widetilde{\bm f}_{10}
& \simeq 1.204981 + 0.26008 \,i\,,
\end{split}
\end{equation}
and
\begin{equation}
\begin{split}
\langle \, \varphi^2 \, \rangle_{+}
& \simeq 1.25976 + 0.43764 \,i\,, \\
\frac{1}{\lambda} +\frac{2}{\sqrt{\lambda}} \, \widetilde{\bm f}_{10} +\widetilde{\bm f}_{20}
& \simeq 1.25981 + 0.43758 \,i\,.
\end{split}
\end{equation}
For the Lefschetz thimble associated with the solution $\varphi = -1/\sqrt{\lambda}$,
we find for $\lambda = 1$ with $N=30$ that
\begin{equation}
\begin{split}
\langle \, \varphi \, \rangle_{-}
& \simeq -1.204933 - 0.26012 \,i\,, \\
{}-\frac{1}{\sqrt{\lambda}}+\widetilde{\bm f}_{10}
& \simeq -1.204981 - 0.26008 \,i\,, 
\end{split}
\end{equation}
and
\begin{equation}
\begin{split}
\langle \, \varphi^{2} \, \rangle_{-}
& \simeq 1.25976 + 0.43764 \,i\,, \\
\frac{1}{\lambda} -\frac{2}{\sqrt{\lambda}} \, \widetilde{\bm f}_{10} +\widetilde{\bm f}_{20}
& \simeq 1.25981 + 0.43758 \,i\,.
\end{split}
\end{equation}

We again found that the correlation functions in the path integral formalism
are reproduced from the formula based on quantum $A_\infty$ algebras
with high precision in all cases,
and the results are consistent with our claim
that the formula gives the correlation functions
on the Lefschetz thimble associated with the solution we chose.
The $n$-point function of the full theory,
\begin{equation}
\langle \, \varphi^{n} \, \rangle
= \frac{Z_{-}}{Z_{-}+Z_{0}+Z_{+}} \, \langle \, \varphi^{n} \, \rangle_{-}
+\frac{Z_{0}}{Z_{-}+Z_{0}+Z_{+}} \, \langle \, \varphi^{n} \, \rangle_{0}
+\frac{Z_{+}}{Z_{-}+Z_{0}+Z_{+}} \, \langle \, \varphi^{n} \, \rangle_{+} \,,
\end{equation}
is reproduced if we know the ratios of the partition functions $Z_{-}$, $Z_{0}$, and $Z_{+}$.

\section{Conclusions and discussion}\label{discussion-section}
\setcounter{equation}{0}

In this paper, we presented a new form of the formula for correlation functions
based on quantum $A_\infty$ algebras
which does not involve
the division of the action into the free part and the interaction part.
For the theory described by the coderivation ${\bf M}$,
the formula associated with a solution $\Phi_\ast$ described
by the coderivation ${\bf \Phi}_\ast$ is given by
\begin{equation}
\langle \, \Phi^{\otimes n} \, \rangle
= \pi_n \, \frac{1}{{\bf I} +i \hbar \, {\bf H} \, {\bf U}} \, {\bf P} \, {\bf 1} \,,
\end{equation}
where ${\bf H}$ satisfies
\begin{equation}
{\bf M} \, {\bf H} +{\bf H} \, {\bf M} = {\bf I} -{\bf P} \,, \qquad 
{\bf H} \, {\bf P} = 0 \,, \qquad
{\bf P} \, {\bf H} = 0 \,, \qquad
{\bf H}^2 = 0
\label{H-conditions-again}
\end{equation}
with
\begin{equation}
{\bf P} = e^{{\bf \Phi}_\ast} \, \pi_0 \,.
\end{equation}
We claim that the formula describes correlation functions
on the Lefschetz thimble associated with the solution $\Phi_\ast$,
and we presented evidence that the formula
contains nonperturbative information on correlation functions
in the case of scalar field theories in zero dimensions.

As we have seen in~\eqref{f_20-Euclidean} and~\eqref{f_40-Euclidean} for the Euclidean case
and in~\eqref{f_20-Lorentzian} and~\eqref{f_40-Lorentzian} for the Lorentzian case,
the $n$-point function ${\bm f}_{n0}$ takes the form of a rational function of the coupling constant
when we truncate the operator
${\bf I} +{\bm h} \, {\bm m} -\hbar \, {\bm h} \, {\bf U}$
or ${\bf I} +{\bm h} \, {\bm m} +i \hbar \, {\bm h} \, {\bf U}$
to a finite matrix,
and this is reminiscent of the Pad\'{e} approximation.
Technically, this will be the reason why we obtained a better approximation than perturbation theory,
but we also emphasize that we did not intend to improve the perturbation theory
and this structure automatically emerged in the process of calculating the inverse of a finite matrix.
Our motivation for the analysis presented in this paper
is to convince ourselves that the formula for correlation functions based on quantum $A_\infty$ algebras
contains information beyond perturbation theory,
and the truncation of ${\bf I} +{\bm h} \, {\bm m} -\hbar \, {\bm h} \, {\bf U}$
or ${\bf I} +{\bm h} \, {\bm m} +i \hbar \, {\bm h} \, {\bf U}$
to a finite matrix was not physically motivated,
but it might be worthwhile to explore a new efficient approximation method
for correlation functions using our approach.

Our construction of ${\bf H}$ is based on the existence of $h$
satisfying $Q \, h +h \, Q = \mathbb{I}$
in the case of a trivial solution
or the existence of $\widetilde{h}$
satisfying $\widetilde{Q} \, \widetilde{h} +\widetilde{h} \, \widetilde{Q} = \mathbb{I}$
in the case of a nontrivial solution.
This does not work in the limit $a \to 0$ of the Airy function
or in the limit $m \to 0$ of $\varphi^4$ theory,
but the path integral converges even in these limits
and there might be a way to construct ${\bf H}$ directly.
For free scalar field theories in higher dimensions,
the existence of $h$
satisfying $Q \, h +h \, Q = \mathbb{I}$
is related to the uniqueness of the vacuum
under an appropriate asymptotic behavior of the scalar field $\varphi$
in the limit $| \varphi | \to \infty$.
In the Lorentzian case, the cohomology of $Q$ is nontrivial
and we have plane wave solutions,
but we choose the asymptotic behavior of the scalar field $\varphi$
such that the imaginary part of $\varphi$ is infinitesimally positive
in the limit ${\rm Re} \, \varphi \to -\infty$
and infinitesimally negative in the limit ${\rm Re} \, \varphi \to \infty$
as in the contour shown in figure~\ref{figure-phi^4-thimble}.
This corresponds to modifying $m^2$ in $Q$ to $m^2 -i \epsilon$ as in~\eqref{S_epsilon},
and the contracting homotopy $h$
satisfying $Q \, h +h \, Q = \mathbb{I}$
can be constructed using the Feynman propagator
so that the perturbation theory in the path integral formalism
is reproduced~\cite{Okawa:2022sjf, Konosu:2023pal, Konosu:2023rkm}.
For a solution corresponding to a nontrivial Lefschetz thimble in higher dimensions,
we expect that the operator ${\bf H}$ satisfying the conditions in~\eqref{H-conditions-again} exists
as the asymptotic behavior of the field
dictated by the Lefschetz thimble gives the unique vacuum,
but we may need to modify $\widetilde{Q}$ for the solution appropriately
when we construct $\widetilde{h}$.
In the analysis of scalar field theories in zero dimensions, however,
we did not need to modify $\widetilde{Q}$
and the structure ${\bf I} +\widetilde{\bm h} \, \widetilde{\bm m}
+ i\hbar \, \widetilde{\bm h} \, {\bf U}$ somehow knows
the asymptotic behavior of the scalar field on the corresponding Lefschetz thimble.
It would be useful to perform concrete analyses in higher dimensions
to see if this is a special feature in zero dimensions.
For a gauge theory,
we need to fix a gauge when we construct $h$ satisfying $Q \, h +h \, Q = \mathbb{I}$.
Once we fix a gauge, our formula may be used even for superstring field theory
under an appropriate identification of $\mathcal{H}_1$ and $\mathcal{H}_2$
as advocated in~\cite{Konosu:2024dpo}.

As we mentioned at the end of subsection~\ref{interpretation-section},
the integration contour $\mathcal{C}$ of the path integral
should be understood as a linear combination of Lefschetz thimbles $\mathcal{J}_i$:
\begin{equation}
\mathcal{C} = \sum_i n_i \, \mathcal{J}_i \,,
\end{equation}
where $i$ labels solutions and $n_i$'s are integers.
While there is a procedure to determine $n_i$ in the path integral formalism~\cite{Witten:2010cx},
we have not found a principle to choose appropriate Lefschetz thimbles
in the language of $A_\infty$ algebras.
Since this is related to unitarity,
it is crucially important to find the corresponding principle
in the description in terms of homotopy algebras.

Furthermore, the correlation functions of the full theory
is given by a linear combination of correlation functions
on Lefschetz thimbles, and contributions from different Lefschetz thimbles
are weighted by partition functions.
We will describe a method to calculate ratios of partition functions
based on quantum $A_\infty$ algebras in a forthcoming paper~\cite{Konosu:2024},
and the correlation functions of the full theory are reproduced
without using the path integral.
However, we have not understood the reason
why we should use partition functions from the viewpoint of homotopy algebras,
and this would be another key question for further development.

Finally, we are hoping that the formula for correlation functions
based on quantum $A_\infty$ algebras in higher dimensions also
contains nonperturbative information.
We can implement the renormalization group
by decomposing the operator ${\bm f}$ into a product of operators
for different energy scales~\cite{Okawa:2024},
and we hope that we can define correlation functions nonperturbatively using this
when there is an ultraviolet fixed point.
Our ambitious goal is then to extend our discussion
to open superstring field theory and define string theory nonperturbatively.

\bigskip

\noindent
{\normalfont \bfseries \large Acknowledgments}

\medskip
We would like to thank Yoshio Kikukawa for helpful discussions on Lefschetz thimbles.
K.~K. would also like to thank Yuji Ando and Jojiro Totsuka-Yoshinaka for discussions.

\end{document}